\documentclass[12pt]{iopart}

\usepackage{graphicx}
\pdfoutput=1
\begin{document}
\title[Measuring information transfer in a soft robotic arm]{Measuring information transfer in a soft robotic arm}
\author{K Nakajima$^{1,2}$, N Schmidt$^{3}$, and R Pfeifer$^{3}$}

\address{$^{1}$ The Hakubi Center for Advanced Research, Kyoto University, 606-8501 Kyoto, Japan \\
$^{2}$ Nonequilibrium Physics and Theoretical Neuroscience, Department of Applied Analysis and Complex Dynamical Systems, Graduate School of Informatics, Kyoto University, 606-8501 Kyoto, Japan \\
$^{3}$ Artificial Intelligence Laboratory, Department of Informatics, University of Zurich, 8050 Zurich, Switzerland}
\ead{jc$\_$mc$\_$datsu@yahoo.co.jp}

\begin{abstract}
Soft robots can exhibit diverse behaviors with simple types of actuation by partially outsourcing control to the morphological and material properties of their soft bodies, which is made possible by the tight coupling between control, body, and environment.
In this paper, we present a method that will quantitatively characterize these diverse spatiotemporal dynamics of a soft body based on the information-theoretic approach. 
In particular, soft bodies have the ability to propagate the effect of actuation through the entire body, with a certain time delay, due to their elasticity. 
Our goal is to capture this delayed interaction in a quantitative manner based on a measure called {\it momentary information transfer}.
We extend this measure to soft robotic applications and demonstrate its power using a physical soft robotic platform inspired by the octopus.  
Our approach is illustrated in two ways.
First, we statistically characterize the delayed actuation propagation through the body as a strength of information transfer.
Second, we capture this information propagation directly as local information dynamics. 
As a result, we show that our approach can successfully characterize the spatiotemporal dynamics of the soft robotic platform, explicitly visualizing how information transfers through the entire body with delays.
Further extension scenarios of our approach are discussed for soft robotic applications in general.
\end{abstract}

\maketitle

\section{Introduction}
Recently, soft materials have been widely used to incorporate flexible elements into robots' bodies. 
These robots are called {\it soft robots} and have significant advantages over traditional articulated robots in terms of morphological deformability and interactional safety \cite{Soft1}. 
They can adapt their morphology to unstructured environments and carry and touch fragile objects without causing damage, which makes them especially applicable for rescue and interaction with humans \cite{Soft2,Soft3}. 
In addition, in \cite{Soft4}, it has been demonstrated that they can generate complex behaviors with simple types of actuation by partially outsourcing control to the morphological and material properties of their soft bodies.
This is enabled by the dynamic coupling between control, body, and environment, which is enhanced by the soft and flexible body \cite{Rolf1,Rolf2,Kohei1}. 

One important difference between conventional rigid robots and soft robots is found in their body dynamics. 
In general, soft body dynamics exhibit diverse properties, such as high-dimensionality, nonlinearity, and a certain amount of ``sluggishness'' with memory, due to their elastic nature \cite{Soft2}.
These properties usually make the robots difficult to control with methods from conventional control theory.
However, in nature, some animals have soft bodies and control them in a sophisticated manner. 
The octopus serves as an extreme example \cite{Oct1}.
It does not have any rigid components in its body, which can execute in virtually infinite degrees of freedom.
Its motion control is far reaching in terms of conventional control framework \cite{Oct3,Oct4}.
Accordingly, roboticists have been investigating how the octopus enables its motion control, searching for a novel control principle for soft robot control \cite{Soft3}. 
For example, a soft robotic arm has been proposed, inspired by the characteristic muscular structure of the octopus called {\it muscular hydrostat}, which enables behaviors such as reaching and grasping \cite{Soft_SSSA3,Soft_SSSA1,Soft_SSSA4}.
Furthermore, exploiting a material characteristic of a soft body, a robot that can crawl toward a target position in an underwater environment by coordinating the arms has been proposed in \cite{Soft_SSSA2,Tao2}.
A novel control scheme inspired by the organization of the octopus's nervous system has been also proposed in \cite{Oct2}, and has been tested and implemented both in a simulator \cite{Kuwabara} and a physical platform \cite{Tao1,Tao3}.
In addition, it has been recently demonstrated that the characteristic structure of the octopus arm can serve as a computational resource, which can be used as a part of a computational device \cite{Kohei2,Kohei3}. 
All of these examples are motivated to positively exploit the material characteristics of the octopus body and the diverse dynamics that the soft body generates.
Based on this context, in this paper, we propose a scheme that effectively and quantitatively characterizes these soft body dynamics.

Our scheme presented in this paper is based on the information-theoretic approach.
This approach is capable of analyzing nonlinear systems, and does not require precise information about the focused system beforehand, which is appropriate when the precise equations of the system are unknown (this is called a model-free approach). 
Due to these properties, it has been actively applied in a wide range of research fields, such as physics, neuroscience \cite{Amblard}, economy \cite{Kantz}, and geological physics \cite{MTE1,MTE2,MTE3}.
In robotics, to characterize the interaction modality of the dynamical coupling between controller, body, and environment, information-theoretic measures, such as {\it mutual information} (MI) and {\it transfer entropy} (TE) \cite{TE}, have been effectively applied \cite{Nico,Max}.
In this paper, we demonstrate that the information-theoretic approach can be successfully applied for soft robots.
In particular, when the soft body is actuated, it can propagate the effect of this actuation through the entire body with a certain amount of time delay. 
We specifically focus on capturing this delayed interaction regime intrinsic to soft body dynamics.
We use a recently proposed measure called {\it momentary information transfer} (MIT) \cite{MTE1}, which is particularly suited to detect delayed couplings among time series, and introduce several variations of this measure to cope with practical requirements when applied to soft robotic platforms. 
Using a physical soft robotic platform, we demonstrate the power of our proposed measures from two aspects.
First, we infer the delayed interaction regime as a strength of information transfer from a statistical point of view.
This aspect is useful for evaluating an intrinsic property corresponding to the soft robotic platform in use.
Second, we monitor and visualize how the information propagates through the body dynamically.
This scheme is beneficial for detecting how external damage in a spatiotemporal point spreads through the entire body, which would be difficult to characterize only by looking at the behavior and the dynamics of the body.

This paper is organized as follows.
In the following section, we present the measure MIT and discuss why this measure can perform better than the conventional TE in inferring delayed information transfer.
Then, we modify this measure systematically to meet specific requirements in real-world applications. 
Here, we propose several variants of MIT, which will be used for our analyses.
Next, we introduce a soft robotic platform inspired by the octopus and explain the experimental procedures to apply these measures.
We demonstrate the effectiveness of our proposed scheme in revealing the delayed information transfer in the soft robotic platform.
Finally, we give concluding remarks, including a possible extension scenario of our approach for general applications to soft robots, and discuss future work.

\section{Methods}
\subsection{Information theoretic approach for delayed information transfer in spatiotemporal dynamics}
\label{sec1}
In this section, we begin with a brief overview of the information-theoretic concepts required in this paper.
The Shannon entropy \cite{Cover} is one of the basic quantities in information theory, which defines the average uncertainty associated with the state $x$ of a random variable $X$ as
\begin{equation}
H(X) = - \sum_{x} p(x) \log_{2} p(x),
\end{equation}
where $p(x)$ denotes a probability distribution of $X$. 
The base of the logarithm is taken as $2$ throughout this paper.
Thus, the unit of all measures presented here is unified as bit. 
The MI between two variables $X$ and $Y$ measures the average information gained about $X$ by knowing $Y$, or vice versa, as follows:
\begin{equation}
M_{XY} = \sum_{x, y} p(x, y) \log \frac{p(x, y)}{p(x) p(y)},
\end{equation}
where $p(x, y)$ is a joint probability distribution of variables $X$ and $Y$ \cite{Cover}.
For statistically independent distributions, $p(x, y) = p(x) p(y)$ and $M_{XY} =0$.
If statistical dependencies exist, $M_{XY} > 0$.
MI is a fundamental measure in information theory and used to evaluate an association between two or more variables, which naturally encompass linear and nonlinear dependencies.
However, as can be seen from the equation, MI is intrinsically symmetric under the exchange of the two variables $X$ and $Y$, which means that MI does not contain any directional information.

In this paper, we focus on the notion of information transfer, which requires capturing both directional and dynamic relations from an information source to a receiver.
Therefore, MI is insufficient for this purpose. 
There are several measures proposed to address these inadequacies. 
A well-known measure is TE, proposed by Schreiber \cite{TE}. 
TE is a measure of the information transfer from the driving system ($Y$) to the responding system ($X$). 
Let us write $x_{t}$ and $y_{t}$ for the values of two temporal processes $X_{t}$ and $Y_{t}$, respectively.
TE essentially quantifies the deviation from the generalized Markov property: 
$p(x_{t+1} | x_{t}^{(K)}) = p(x_{t+1} | x_{t}^{(K)}, y_{t}^{(L)})$, where $p(x_{t+1} | x_{t}^{(K)})$ denotes a transition probability from $x_{t}^{(K)}$ to $x_{t+1}$, and $K$ and $L$ are the length of the embedding vectors.
If the deviation from a generalized Markov process is small, then the state $y_{t}^{(L)}$ can be assumed to have little relevance to the transition probabilities from $x_{t}^{(K)}$ to $x_{t+1}$. 
If the deviation is large, however, then the assumption of a Markov process is not valid. 
The incorrectness of the assumption can be expressed by the TE, formulated as a specific version of the Kullback-Leibler entropy \cite{TE,Cover}:
\begin{equation}\label{TE}
TE_{Y \rightarrow X} = \sum_{x_{t+1}, x_{t}^{(K)}, y_{t}^{(L)}} p(x_{t+1}, x_{t}^{(K)}, y_{t}^{(L)}) \log \frac{p(x_{t+1} | x_{t}^{(K)}, y_{t}^{(L)})}{p(x_{t+1} | x_{t}^{(K)})},
\end{equation}
where the index $TE_{Y \rightarrow X}$ indicates the influence of $Y$ on $X$ and can thus be used to detect the directed information transfer. 
In other words, TE measures how well we can predict the transition of system $X$ by knowing system $Y$. 
TE is non-negative, and any information transfer between the two variables results in $TE_{Y \rightarrow X}>0$. 
If the state $y_{t}^{(L)}$ has no influence on the transition probabilities from $x_{t}^{(K)}$ to $x_{t+1}$, or if the two time series are completely synchronized, then $TE_{Y \rightarrow X}=0$.
By introducing a time delay $\tau$, TE has been used to infer delayed couplings such as
\begin{equation}\label{TE}
TE_{Y \rightarrow X} (\tau) = \sum_{x_{t+\tau}, x_{t}^{(K)}, y_{t}^{(L)}} p(x_{t+\tau}, x_{t}^{(K)}, y_{t}^{(L)}) \log \frac{p(x_{t+\tau} | x_{t}^{(K)}, y_{t}^{(L)})}{p(x_{t+\tau} | x_{t}^{(K)})},
\end{equation}
where the index $TE_{Y \rightarrow X}$ indicates the influence of $Y$ on $X$ with a delay $\tau$ ($\tau = 1, 2, ...$).

A weak point of TE is that it fails at detecting delayed couplings \cite{MTE1,Wibral}.
One attempt to overcome this drawback, and to capture delayed information transfer, has been presented in \cite{MTE1}, where the authors proposed the measure MIT, defined as
\begin{equation}\label{MIT}
MIT_{Y \rightarrow X} ( \tau ) = \sum_{x_{t+\tau}^{(K+1)}, y_{t}^{(L)}} p(x_{t+\tau}^{(K+1)}, y_{t}^{(L)}) \log \frac{p(x_{t+\tau} | x_{t+\tau-1}^{(K)}, y_{t}^{(L)})}{p(x_{t+\tau} | x_{t+\tau-1}^{(K)}, y_{t-1}^{(L-1)})},
\end{equation}
where the index $MIT_{Y \rightarrow X} ( \tau )$ indicates the MIT from $y_{t}$ to $x_{t+\tau}$ and $\tau$ ($\tau = 1, 2, ...$) denotes a time delay.
Note that, in the equation, $(x_{t+\tau}, x_{t+\tau-1}^{(K)})$ and $(y_{t}, y_{t-1}^{(L-1)})$ are unified into $x_{t+\tau}^{(K+1)}$ and $y_{t}^{(L)}$, respectively, for simplicity.
It is obvious from the equation that MIT represents the information transfer from $y_{t}$ to $x_{t+\tau}$ under the condition of the joint past of $y_{t}$ and $x_{t+\tau}$. 
This specific type of conditioning enables MIT to have a better resolution for detecting point-to-point information transfer in a temporal dimension.
Similar considerations have also been presented in \cite{Polani,Lizier2}.
Further details on the performance of MIT in a real-world time series, and on the comparisons with TE are given in \cite{MTE1}.

Our aim in this paper is to apply information-theoretic measures to spatiotemporal time series in a physical soft robot platform.
In particular, we are interested in capturing the delayed information transfer through soft body dynamics and characterizing how it propagates through the entire body.
A number of useful methods for this purpose have been proposed in the literature in different contexts.
Based on MIT, we will systematically place specific conditions for our requirements and explain how to incorporate these methods and obtain feasible measures for our purpose.
This will help us to understand how each extension clearly meets our requirement and assist in applying our approach to any soft robotic platform in use. 
A rough sketch of our requirements and modification strategies of the measure are given next, which will be explained in detail in the following sections.

The first requirement is for ease in handling the measure for a real-world time series.
As previously explained, the information-theoretic measures are functionals of probability distributions, and there are variations to methods proposed to estimate the distributions from an obtained time series.
In a real-world time series, this procedure often requires careful preprocessing of the data, setting of the distribution estimators, and fine-tuning their parameters, due to environmental and observational noises. 
Furthermore, the obtained results are usually not easy to reproduce without specifying the details of these methods.
One of the well-known treatments for these issues is to introduce a class of the probability distribution estimator based on permutation partitioning for the values of the data.
The corresponding uncertainty measure, {\it permutation entropy} (PE), is introduced in \cite{PE1} and quantifies the uncertainty of the local orderings of values, unlike the usual entropy, which quantifies the uncertainty of the values themselves. 
This approach enables a natural discretization for time series data that does not require any knowledge about the range of values of the time series beforehand.
It can be applied to raw data without further model assumptions, because the permutation orderings refer only the local and neighboring values in time series data.
The information-theoretic measures based on the permutation partitioning have been demonstrated to be easy to implement relative to the other traditional methods \cite{PE1}, fast to compute \cite{STE}, robust against noise \cite{PE1,PE3,PE5,TERV2,Kohei4}, and applicable to non-stationary and transient time series data \cite{PE4}, which meets our requirement.

The second requirement is to make the measure applicable to a multivariate time series.
Body dynamics generated by soft robotic platforms often reveal high-dimensional spatiotemporal dynamics compared with rigid ones.
As we saw for MIT, measuring delayed couplings in high resolution needs special care in conditioning the past time series, both for the information source and receiver.
We will explain how to deal with this conditioning for multivariate time series data in detail, specifically for data obtained from the physical platform.

The final requirement is to make it possible for the measure to characterize a local information transfer profile within the spatiotemporal dynamics.
It is valuable to track and visualize how the information dynamically propagates through the entire body, along with its corresponding body dynamics.
However, just like other information-theoretic measures, due to their statistical nature, MIT only provides an expectation value of the amount of information transferred that originates from the global average, which does not directly correspond to the body dynamics themselves.
We will explain in detail, a method to make MIT applicable for our purpose by localizing it to each spatiotemporal point of the body dynamics. 

\subsubsection{Momentary sorting information transfer: Permutation analogue of MIT}
As previously explained, PE quantifies the uncertainty of the local orderings of values, instead of the uncertainty of the values themselves.
Let $x_{t}^{(L)}$ represent an $L$-dimensional embedding vector from the obtained time series $x'_{t}$, and $\hat{x}_{t}^{(L)}$ be a sequence of numbers representing the orderings of $x_{t}^{(L)}$. 
Based on the permutations of the values, $\hat{x}_{t}^{(L)}$ is generated as follows: 
$x_{t}^{(L)} = ( x'_{t}, x'_{t-1}, ..., x'_{t-(L-1)} )$, and the values are arranged in ascending order, 
$ x'_{t-(o_{t} (1) -1)} \leq  x'_{t-(o_{t} (2) -1)} \leq ... \leq x'_{t-( o_{t} (L) -1)} $.
A symbol is thus defined as $\hat{x}_{t}^{(L)} \equiv (o_{t} (1), o_{t} (2), ..., o_{t} (L) ) \in \hat{X}_{t}$, where $\hat{X}_{t}$ is the set of symbols generated in the temporal process $X_{t}$.
Based on the generated symbols $\hat{x}_{t}^{(L)}$, PE is expressed as:
\begin{equation}
H(\hat{X}_{t}) = - \sum_{\hat{x}_{t}^{(L)}} p(\hat{x}_{t}^{(L)}) \log p(\hat{x}_{t}^{(L)}),
\end{equation}
where $p(\hat{x}_{t}^{(L)})$ is the probability of the occurrence of $\hat{x}_{t}^{(L)}$ in the set of symbols $\hat{X}_{t}$.
In spite of the differences between the procedures, it was proven that the PE rate is equal to the usual entropy rate for some conditions \cite{PE2,Amigo_book,Amigo1,Amigo3,Haruna1}. 

A number of permutation analogues for the information-theoretic measures have been proposed and the relations with their original measure have been investigated (e.g., systematic investigations on Kolmogorov-Sinai entropy can be found in \cite{Amigo2,Keller1,Keller2,Keller3,Keller4,Keller5} and investigations on TE and its related measures can be found in \cite{Haruna2,Haruna3,Haruna4}).
Similar to MIT, its permutation version, proposed in \cite{MTE1}, is called {\it momentary sorting information transfer} (MSIT), expressed as:
\begin{equation}\label{MIT}
MSIT_{Y \rightarrow X} ( \tau ) = \sum_{\hat{x}_{t+\tau}^{(K+1)}, \hat{y}_{t}^{(L)}} p(\hat{x}_{t+\tau}^{(K+1)}, \hat{y}_{t}^{(L)}) \log \frac{p(\hat{x}_{t+\tau} | \hat{x}_{t+\tau-1}^{(K)}, \hat{y}_{t}^{(L)})}{p(\hat{x}_{t+\tau} | \hat{x}_{t+\tau-1}^{(K)}, \hat{y}_{t-1}^{(L-1)})},
\end{equation}
where, similar with MIT, the index $MSIT_{Y \rightarrow X} ( \tau )$ indicates the MSIT of $\hat{y}_{t}$ to $\hat{x}_{t+\tau}$ and $\tau$ ($\tau = 1, 2, ...$) denotes a time delay.
Note that for $\hat{x}_{t+\tau}$, the permutation orderings among $\hat{x}_{t+\tau}^{(K+1)}$ are considered, because when we calculate the measure, we first obtain the joint entropies from the corresponding probability distributions, namely $p(\hat{x}_{t+\tau}^{(K+1)}, \hat{y}_{t}^{(L)})$ and $p(\hat{x}_{t+\tau}^{(K+1)}, \hat{y}_{t-1}^{(L-1)})$, with the obtained time series.
A similar treatment has been also adopted in \cite{TERV1}.

In the next section, we will show how MSIT can be extended for the case of multivariate time series.

\subsubsection{MSIT for multivariate time series}
As we saw in the case for M(S)IT, it is important to employ a conditioning for the information receiver and sender in focus.
This guarantees to yield a high-resolution measure for the bivariate time series.
When we extend the measure for the multivariate case, we have to take particular care of the conditioning, because other possible causal information contributors---apart from the information source under consideration---may be involved in the time series as well.
It is therefore important to ``filter out'' the effect of these other information sources, which might otherwise occlude effect of the source in question.
There have been several studies to address these issues.
In \cite{MTE2,MTE3}, the generalized MIT has been proposed, a method to infer all causal contributors to the receiver, based on the graphical model approach.
Their approach is general, because the method allows reconstruction of an interaction modality of the multivariate data in terms of a process graph.
This graph is built from scratch without having any knowledge about the dynamical property of the underlying system in focus (such as network topology of the system).

\begin{figure}[htbp]
\centerline{\includegraphics[width=3.0in, bb=18 314 312 744, clip]{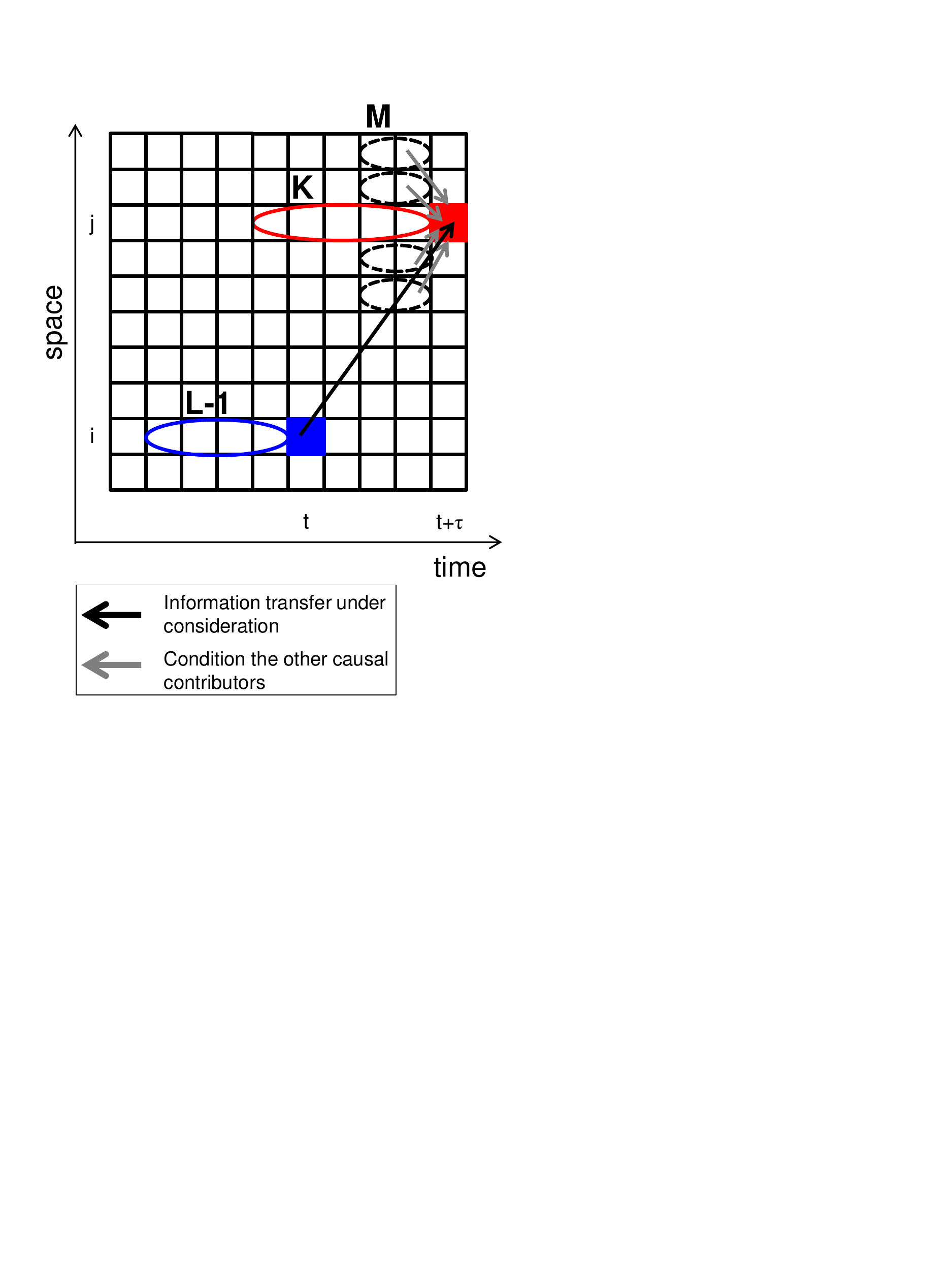}}%MSIT_space2.eps
\caption{Schematic expressing the MSIT in a spatiotemporal system. The black arrow shows the information transfer under consideration and the gray arrow shows the other causal contributors to the destination cell. The MSIT for the case when $(\tau, K, L, M, N_{r}, T_{r}) = (4, 5, 5, 2, 2, 1)$ is shown.}
\label{MSIT_space}
\end{figure}

In this paper, we will adopt their approach only partially.
We take into account a physical constraint of the soft robotic platform, as a support for estimating the other causal information contributors, apart from the information source.
We first assume that the experimenter knows where each time series comes from in space, namely the location in the physical body.
We further assume that the degree of information transfer depends on the relative spatial and temporal distance; if the information source and receiver are more distant and delayed, then the strength of the information transfer decreases, which would be a natural assumption, especially when the information transfer substrates are mediated with physical materials. 
Based on these assumptions, we introduce a simple one-dimensional space as an example, where each point in space is expressed as cell $i$ ($i=1, 2, ..., N$), resulting in $N$ time series.
Accordingly, the senders and receivers are denoted as spatially ordered cells, the MSIT for spatiotemporal time series from cell $i$ to cell $j$ can be expressed as
\begin{equation}\label{MSIT}
\fl MSIT_{i \rightarrow j} ( \tau ) = \sum_{\hat{x}_{j, t+\tau}^{(K+1)}, \hat{x}_{i, t}^{(L)}, \hat{v}_{i, j, t}} p(\hat{x}_{j, t+\tau}^{(K+1)}, \hat{x}_{i, t}^{(L)}, \hat{v}_{i, j, t}) \log \frac{p(\hat{x}_{j, t+\tau} | \hat{x}_{j, t+\tau-1}^{(K)}, \hat{x}_{i, t}^{(L)}, \hat{v}_{i, j, t})}{p(\hat{x}_{j, t+\tau} | \hat{x}_{j, t+\tau-1}^{(K)}, \hat{x}_{i, t-1}^{(L-1)}, \hat{v}_{i, j, t})},
\end{equation}
\begin{eqnarray}\label{condition}
\fl \hat{v}_{i, j, t} = \{ \hat{x}_{q, t+\tau'}^{(M)} | \forall (q, \tau'), 1 \leq q \leq N, \tau' < \tau \ (except \ \hat{x}_{j, t+\tau}^{(K+1)}, \hat{x}_{i, t}^{(L)}) \},
\end{eqnarray}    
where $\hat{v}_{i, j, t}$ denotes possible information sources to cell $j$ excluding $\hat{x}_{j, t+\tau}^{(K+1)}$ and $\hat{x}_{i, t}^{(L)}$, and $M, q,$ and $\tau'$ denote their embedding dimension, cell number, and a time delay, respectively.
One could adjust these parameters for each information source; for example, if there are overlapping sources in a temporal dimension, such as $(\hat{x}_{q, t+1}^{(M)}, \hat{x}_{q, t}^{(M)})$, we can combine them into one as $\hat{x}_{q, t+1}^{(M+1)}$.
Note that it is possible to condition information sources for $\hat{x}_{i, t}^{(L)}$ to sharpen the point-to-point MIT, but in this paper, we only consider $\hat{x}_{i, t}$ under the condition of its direct past $\hat{x}_{i, t-1}^{(L-1)}$ for simplicity. 
Now, we specify the conditions of Equations (\ref{MSIT}) and (\ref{condition}) for our spatiotemporal constraint, expressed as
\begin{equation}\label{MSIT_ST}
\fl MSIT^{ST}_{i \rightarrow j} ( \tau ) = \sum_{\hat{x}_{j, t+\tau}^{(K+1)}, \hat{x}_{i, t}^{(L)}, \hat{v}^{ST}_{i, j, t}} p(\hat{x}_{j, t+\tau}^{(K+1)}, \hat{x}_{i, t}^{(L)}, \hat{v}^{ST}_{i, j, t}) \log \frac{p(\hat{x}_{j, t+\tau} | \hat{x}_{j, t+\tau-1}^{(K)}, \hat{x}_{i, t}^{(L)}, \hat{v}^{ST}_{i, j, t})}{p(\hat{x}_{j, t+\tau} | \hat{x}_{j, t+\tau-1}^{(K)}, \hat{x}_{i, t-1}^{(L-1)}, \hat{v}^{ST}_{i, j, t})},
\end{equation}
\begin{eqnarray}\label{condition_ST}
\fl \hat{v}^{ST}_{i, j, t} = \{ \hat{x}_{j \pm r, t+\tau-\tau_{r}}^{(M)} | \forall (r, \tau_{r}), 1 \leq r \leq N_{r}, 1 \leq \tau_{r} \leq T_{r} \ (except \ \hat{x}_{j, t+\tau}^{(K+1)}, \hat{x}_{i, t}^{(L)}) \},
\end{eqnarray}
where $MSIT^{ST}_{i \rightarrow j} ( \tau )$ and $\hat{v}^{ST}_{i, j, t}$ represent the MIT from $i$ to $j$ with delay $\tau$ and the possible information sources to cell $j$ under the spatiotemporal constraint, respectively, parameters $r$ and $\tau_{r}$ determine a cell and a time delay for possible information sources to cell $j$ excluding $\hat{x}_{j, t+\tau}^{(K+1)}$ and $\hat{x}_{i, t}^{(L)}$, and $N_{r}$ and $T_{r}$ specify the spatial and temporal range around the information receiver, respectively; if $r$ and $\tau_{r}$ get larger, then the possible information source is more spatiotemporally distant from the receiver. 
Figure \ref{MSIT_space} visualizes the $MSIT^{ST}_{i \rightarrow j} ( \tau )$ and summarizes the effects of the different parameters involved. 

Note that $MSIT^{ST}_{i \rightarrow j} ( \tau )$ can be also denoted as a sum of permutation joint entropies:
\begin{eqnarray}
\fl MSIT^{ST}_{i \rightarrow j} ( \tau ) = -H(\hat{X}_{t+\tau}^{j}, \hat{X}_{t}^{i}, \hat{V}_{t}^{i, j}) +H(\hat{X}_{t+\tau-1}^{j}, \hat{X}_{t}^{i}, \hat{V}_{t}^{i, j}) \\
\ \ \ \ \ \ \ \ \ \ \ \ \ \ \ \ \ \ \ \ \ \ \ -H(\hat{X}_{t+\tau}^{j}, \hat{X}_{t-1}^{i}, \hat{V}_{t}^{i, j}) +H(\hat{X}_{t+\tau-1}^{j}, \hat{X}_{t-1}^{i}, \hat{V}_{t}^{i, j}),
\end{eqnarray}
where $\hat{X}_{t+\tau}^{j}$, $\hat{X}_{t+\tau-1}^{j}$, $\hat{X}_{t}^{i}$, $\hat{X}_{t-1}^{i}$, and $\hat{V}_{t}^{i, j}$ are the temporal processes for the values of $\hat{x}_{j, t+\tau}^{(K+1)}$, $\hat{x}_{j, t+\tau-1}^{(K+1)}$, $\hat{x}_{i, t}^{(L)}$, $\hat{x}_{i, t-1}^{(L)}$, and $v^{ST}_{i, j, t}$, respectively, and $H(\hat{X}_{t+\tau}^{j}, \hat{X}_{t}^{i}, \hat{V}_{t}^{i, j})$, $H(\hat{X}_{t+\tau-1}^{j}, \hat{X}_{t}^{i}, \hat{V}_{t}^{i, j})$, $H(\hat{X}_{t+\tau}^{j}, \hat{X}_{t-1}^{i}, \hat{V}_{t}^{i, j})$, and $H(\hat{X}_{t+\tau-1}^{j}, \hat{X}_{t-1}^{i}, \hat{V}_{t}^{i, j})$ are joint entropies of corresponding temporal processes.
When we calculate the value of $MSIT^{ST}_{i \rightarrow j} ( \tau )$, we first calculate the joint and single entropies on the right side of this equation from the corresponding probability distributions, namely $p(\hat{x}_{j, t+\tau}^{(K+1)}, \hat{x}_{i, t}^{(L)}, \hat{v}^{ST}_{i, j, t})$, $p(\hat{x}_{j, t+\tau-1}^{(K+1)}, \hat{x}_{i, t}^{(L)}, \hat{v}^{ST}_{i, j, t})$, $p(\hat{x}_{j, t+\tau}^{(K+1)}, \hat{x}_{i, t-1}^{(L)}, \hat{v}^{ST}_{i, j, t})$, and $p(\hat{x}_{j, t+\tau-1}^{(K+1)}, \hat{x}_{i, t-1}^{(L)}, \hat{v}^{ST}_{i, j, t})$, with the obtained time series.

\subsubsection{Localizing MSIT for spatiotemporal systems}
\label{sec3}
In this section, we present the procedure for making MSIT local.
This procedure makes it possible to reveal an information transfer profile for each spatiotemporal point by directly corresponding the measure to the observed time series, which would be especially useful for monitoring the interaction modality of the local dynamics from the information-theoretic point of view. 
We keep carrying on the physical constraint introduced in the previous section and explain the procedure based on the measure in Equations (\ref{MSIT_ST}) and (\ref{condition_ST}).

In \cite{Lizier1}, Lizier {\it et al.} focused on the fact that, in calculating the information-theoretic measure from the experimental data, the associated probability $p(x)$ is operationally equivalent to the ratio of the count of observations $c(x)$ of states, to the total number of observations $O$ made. 
In applications to time series data for MSIT, the number of observations is finite, and $p(\hat{x}_{j, t+\tau}^{(K+1)}, \hat{x}_{i, t}^{(L)}, \hat{v}^{ST}_{i, j, t})$ in Equation (\ref{MSIT_ST}) can be expressed as $p(\hat{x}_{j, t+\tau}^{(K+1)}, \hat{x}_{i, t}^{(L)}, \hat{v}^{ST}_{i, j, t}) = c(\hat{x}_{j, t+\tau}^{(K+1)}, \hat{x}_{i, t}^{(L)}, \hat{v}^{ST}_{i, j, t})/O$. 
Then, MSIT in Equation (\ref{MSIT_ST}) can be expressed as follows:
\begin{equation}
\fl
MSIT^{ST}_{i \rightarrow j} ( \tau ) = \frac{1}{O} \sum_{\hat{x}_{j, t+\tau}^{(K+1)}, \hat{x}_{i, t}^{(L)}, \hat{v}^{ST}_{i, j, t}} \left( \sum_{a=1}^{c(\hat{x}_{j, t+\tau}^{(K+1)}, \hat{x}_{i, t}^{(L)}, \hat{v}^{ST}_{i, j, t})} 1 \right) \log \frac{p(\hat{x}_{j, t+\tau} | \hat{x}_{j, t+\tau-1}^{(K)}, \hat{x}_{i, t}^{(L)}, \hat{v}^{ST}_{i, j, t})}{p(\hat{x}_{j, t+\tau} | \hat{x}_{j, t+\tau-1}^{(K)}, \hat{x}_{i, t-1}^{(L-1)}, \hat{v}^{ST}_{i, j, t})}.
\end{equation}
By considering that a double sum running over each actual observation $a$ for each possible tuple observation $(\hat{x}_{j, t+\tau}^{(K+1)}, \hat{x}_{i, t}^{(L)}, \hat{v}^{ST}_{i, j, t})$ is nothing but a single sum over all $O$ observations, we obtain the following:
\begin{equation}
MSIT^{ST}_{i \rightarrow j} ( \tau ) = \frac{1}{O} \sum_{all \ observations} \log \frac{p(\hat{x}_{j, t+\tau} | \hat{x}_{j, t+\tau-1}^{(K)}, \hat{x}_{i, t}^{(L)}, \hat{v}^{ST}_{i, j, t})}{p(\hat{x}_{j, t+\tau} | \hat{x}_{j, t+\tau-1}^{(K)}, \hat{x}_{i, t-1}^{(L-1)}, \hat{v}^{ST}_{i, j, t})}.
\end{equation}
Thus, we can write MSIT as the global average over {\it local momentary sorting information transfer} (LMSIT), $LMSIT^{ST}_{i \rightarrow j} (\tau)$, defined as,
\begin{equation}
MSIT^{ST}_{i \rightarrow j} ( \tau ) =\langle LMSIT^{ST}_{i \rightarrow j} (\tau) \rangle,
\end{equation}
\begin{equation}
LMSIT^{ST}_{i \rightarrow j} (\tau) = \log \frac{p(\hat{x}_{j, t+\tau} | \hat{x}_{j, t+\tau-1}^{(K)}, \hat{x}_{i, t}^{(L)}, \hat{v}^{ST}_{i, j, t})}{p(\hat{x}_{j, t+\tau} | \hat{x}_{j, t+\tau-1}^{(K)}, \hat{x}_{i, t-1}^{(L-1)}, \hat{v}^{ST}_{i, j, t})},
\end{equation}
where $\langle X \rangle$ denotes the expectation value of $X$.
The term $\hat{v}^{ST}_{i, j, t}$ takes the same expression as Equation (\ref{condition_ST}).
Note that LMSIT can have a negative value. 
The negative value of LMSIT means that the sender is misleading about the prediction of the receiver's next state \cite{Lizier1}.
This measure is defined for every spatiotemporal receiver $(j, t+\tau)$, forming a spatiotemporal profile for every information source $i$ in focus.
Thus, in a practical sense, as we see later, it would be more reasonable to adjust the time index for the information receiver as $(j, t)$.
This can be easily done by simply shifting the time index without altering the important factor of the measure as
\begin{equation}\label{LMSIT_final}
LMSIT^{ST}_{i \rightarrow j} (\tau) = \log \frac{p(\hat{x}_{j, t} | \hat{x}_{j, t-1}^{(K)}, \hat{x}_{i, t-\tau}^{(L)}, \hat{v}^{ST}_{i, j, t-\tau})}{p(\hat{x}_{j, t} | \hat{x}_{j, t-1}^{(K)}, \hat{x}_{i, t-\tau-1}^{(L-1)}, \hat{v}^{ST}_{i, j, t-\tau})},
\end{equation}
\begin{eqnarray}\label{LMSIT_condition_final}
\fl \hat{v}^{ST}_{i, j, t-\tau} = \{ \hat{x}_{j \pm r, t-\tau_{r}}^{(M)} | \forall (r, \tau_{r}), 1 \leq r \leq N_{r}, 1 \leq \tau_{r} \leq T_{r} \ (except \ \hat{x}_{j, t}^{(K+1)}, \hat{x}_{i, t-\tau}^{(L)}) \}.
\end{eqnarray}
This forms the expression of LMSIT addressed throughout this paper.
Similar local measures have been proposed in our earlier works \cite{Kohei4,Robio2}, which are based on the permutation version of TE called {\it symbolic transfer entropy} (STE) \cite{STE}.
Our proposed measures extend our previous measures in terms of the condition of the information sender, which is expected to sharpen resolution, especially for detecting delayed information transfer.
As in the case with $MSIT^{ST}_{i \rightarrow j} ( \tau )$, to calculate the value of $LMSIT^{ST}_{i \rightarrow j} (\tau)$, we calculate the joint and single probability distributions, namely $p(\hat{x}_{j, t}^{(K+1)}, \hat{x}_{i, t-\tau}^{(L)}, \hat{v}^{ST}_{i, j, t-\tau})$, $p(\hat{x}_{j, t-1}^{(K+1)}, \hat{x}_{i, t-\tau}^{(L)}, \hat{v}^{ST}_{i, j, t-\tau})$, $p(\hat{x}_{j, t}^{(K+1)}, \hat{x}_{i, t-\tau-1}^{(L)}, \hat{v}^{ST}_{i, j, t-\tau})$, and $p(\hat{x}_{j, t-1}^{(K+1)}, \hat{x}_{i, t-\tau-1}^{(L)}, \hat{v}^{ST}_{i, j, t-\tau})$, with the obtained time series.

In this section, we introduced MSIT and LMSIT with spatiotemporal constraints to meet our requirements.
In the next section, we will explain our physical platform with soft robotic arm, and by using the body dynamics generated by the arm, we illustrate how these measures can capture the characteristic properties of soft body dynamics, in particular, the delayed information transfer, and demonstrate the power of the measures.

\subsection{Experimental setup}
In this section, our experimental platform, equipped with a soft robotic arm, is presented and the experimental procedures are explained.
\subsubsection{Platform}
\begin{figure}[htbp]
\centerline{\includegraphics[width=5.5in,bb=30 464 521 774, clip]{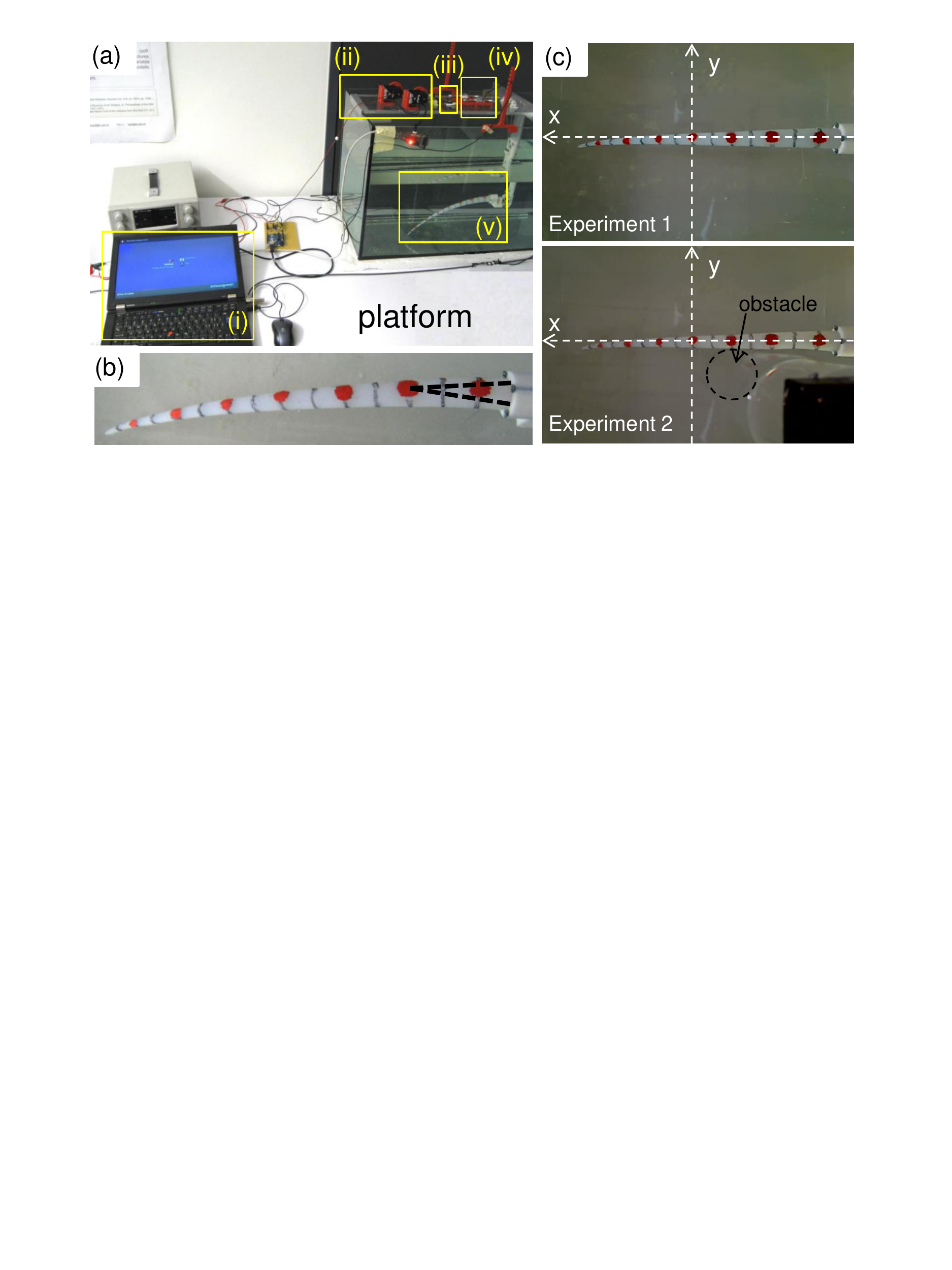}} %platform.eps
\caption{Experimental setups. (a) An overall experimental platform. It consists of a laptop PC (i), two servo motors (ii), one video camera (iii), two force sensors (iv), and a soft robotic arm (v). (b) Close-up image of the soft robotic arm used in this paper. Dashed lines represent the cables embedded in the arm. Red dots and black lines on the arm are references for visual tracking. (c) Initial arm configurations of Experiment 1 (upper figure) and Experiment 2 with the obstacle (lower figure). In each experiment, the arm is set to the relaxed position. In Experiment 2 with the obstacle, the obstacle is made of a transparent plastic tube on a black mount and is set near the R3 of the arm to affect its behavior. The obstacle is fixed so that the arm motion cannot change the position of the obstacle. The base point of the x-y coordinates for the tracking is set to near R4 of the arm in both experiments.} \label{platform}
\end{figure}

The experimental platform consists of a soft robotic arm, its actuation, sensing, control systems, and a water tank containing fresh water as the underwater environment (Fig. \ref{platform} (a)).
There are several soft robotic arms proposed in the literature (see, e.g., \cite{Soft_SSSA1,Soft_SSSA2,Soft_SSSA3,Soft_SSSA4}).
Our soft robotic arm is based on the arm proposed in \cite{Soft_SSSA1}, which mimics the morphology and material characteristics of the octopus arm, and is made of commercially available silicone rubber (ECOFLEX\texttrademark 00-30). 
The arm embeds two nonextensible fishing cables symmetrically to the center of the arm, as shown by the dashed lines in Fig. \ref{platform} (b).
By pulling these two cables via servo motors (Dynamixel\texttrademark AX-12A+), the arm is actuated (Fig. \ref{platform} (a)).
The total length of the cone-shaped soft arm is 310 mm.
The active part of the arm is 80 mm and the remaining 230 mm part of the arm is completely passive.
Two force sensors measure the tension of the cables.
The force sensor signals are amplified and sent to a PC serial port through an Arduino\texttrademark UNO board, with ADC outputs integer values between 0 and 1024, which correspond linearly to forces of 0 to 10 N (Fig. \ref{platform} (a)).
A camera (Logitech\texttrademark Webcame Pro 9000) is placed on the top of the platform to record the motion of the soft silicone arm (Fig. \ref{platform} (a)).

A java program running on a laptop PC sends the motor commands to the motors.
The unit representing ``time'' in this paper is a ``timestep'' based on the sensing and actuation loop in the program (this is about 0.03 s in a physical time). 
For each timestep, a motor command is sent to each motor and the cable tensions are recorded.
According to the motor command, the servo motor angle is adjusted. 
In this paper, motor commands are set as binary values, $M = \{+1, -1 \}$. 
If the command gives $+1$ or $-1$, then the motor is controlled to move from the current position toward the maximum position ($L_{max}$) or the relaxed position ($L_{relax}$), respectively. 
For each motor, $L_{max}$ was determined to avoid the tip of the arm touching the water tank while moving. 
Note that, in the current setting, the motor command does not always take the roller position to $L_{max}$ or $L_{relax}$, but rather decides the direction in which to move for each timestep. 
If the command gives $+1$ or $-1$ when the current position is in $L_{max}$ or $L_{relax}$, respectively, then the position will not change from the current position.

For the analyses, we collected the time series of the two motor commands (M1 and M2) and two force sensors (f1 and f2) for each timestep and tracked the positions of the reference points on the arm (R) in the x-y plane, by using the tracker proposed in \cite{Tracker} (Fig. \ref{platform} (b) and Fig. \ref{timeseries_preparation}).
The tracked time series were manually aligned to the motor command time series after each run.
For the analyses in this paper, we used only 6 reference points, which are indexed from R1 to R6, from the base to the tip.
The nearest reference point to the tip was too fast to allow stable tracking of the motion and was not used for the analyses.
Based on these 6 reference points for each timestep, we fitted the points with a 5th order polynomial curve and for each interval between two neighboring reference points, we segmented the curve into 20 equidistant fragments, resulting in 101 endpoints for fragments, including the reference points (Fig. \ref{timeseries_preparation}, and see detailed procedures in Appendix).
The dynamics of these 101 points are used as a representation of the arm's body dynamics (S), which are indexed from $S_{1}$ to $S_{101}$ from the base to the tip.
For the body dynamics S, we used only the y-coordinate for the information-theoretic analysis, which reflects the motion of the arm better than the x-coordinate (Fig. \ref{platform} (c)).

\subsubsection{Experimental procedures}
We conducted two groups of experiments, where all the experiments were based on open-loop control.
Experiment 1 was designed to reveal the informational structure intrinsic to the soft robotic arm in our platform setting. 
As explained in the previous section, the arm is controlled by two cables embedded in the arm. 
First, we investigated how the effects of these actuations transfer through the entire body. 
Second, we investigated how each body part transfers the information to the other parts of the body.
These two investigations employ the MSIT, which measures delayed couplings.
In this experiment, to avoid correlations induced by coordinated motor commands, we sent random motor commands to each motor for each timestep. 
By doing this, we moved each roller in random positions from $L_{relax}$ to $L_{max}$ over time, and effectively detect the coupling regime intrinsic to the setting. 
We ran the experiment for 5 trials with 5000 timesteps each, starting from the same initial arm configuration (Fig. \ref{platform} (c)).

In Experiment 2, we investigated a local information transfer profile in the spatiotemporal dynamics of our soft robotic arm.
We set a simple oscillatory motor command and observed the behavior with an obstacle in the water tank (Fig. \ref{platform} (c)). 
In this experiment, we aimed to capture how the environmental impact spreads dynamically through the entire body.
This is difficult to reveal directly with a measure that deals with a global average of information transfer, because the effect of damage or contact to the body is usually not a constant event, but rather one that occurs at a specific point of time and space.
We placed a round plastic tube (obstacle) next to the arm, which was set to affect the behavior of the arm (Fig. \ref{platform} (c)). 
For the motor commands, +1 and -1 were set for T timesteps in alternate shifts in an opposite phase each. 
The time window T was set to the maximum timesteps needed for the motor to rotate from the relaxed position to the maximum one, which, in our setting, was 10 timesteps. 
Accordingly, one cycle resulted in 20 timesteps.
We ran the system for 5000 timesteps, which were 250 cycles, from the initial arm configuration depicted in Fig. \ref{platform} (c).
In the experiment, we also ran the system with random motor commands and the same oscillatory motor command without the obstacle and compared the results.

Note that, for both experiments, the experimental condition is designed to make the time series data stationary, and the probability distributions of the variables required to calculate measures, MSIT and LMSIT, are estimated via relative frequencies obtained from the data.
Accordingly, in each experimental trial for both experiments, first 100 timesteps are discarded to avoid the initial transients in the time series data for the analyses.
This procedure is used for all the experimental trials throughout this paper.

Preliminary investigations of information-theoretic analyses on a soft robotic platform have been performed in \cite{Robio2,Robio1} without considering the effect of delayed information transfer. 
In these works, the analyses were conducted based on STE.
However, as explained earlier, STE has drawbacks in detecting delayed information transfer. 
In this paper, by introducing MSIT, we have largely extended our previous approach.

\section{Results}
In this section we present the results for both Experiment 1 and 2 in detail.
As explained in the previous section, Experiment 1 is focused on revealing the delayed information transfer intrinsic to the soft robotic arm.
For the analysis of this experiment, we make use of MSIT expressed in Equation (\ref{MSIT_ST}) and (\ref{condition_ST}) (that is, $MSIT^{ST}_{i \rightarrow j} ( \tau )$).
In Experiment 2, unlike the analysis in Experiment 1, we aim to reveal how the information propagates dynamically through the body by visualizing the local information transfer in each spatiotemporal point.
This analysis is conducted by using LMSIT expressed in Equation (\ref{LMSIT_final}) and (\ref{LMSIT_condition_final}) (that is, $LMSIT^{ST}_{i \rightarrow j} ( \tau )$).  
Throughout our analyses, the embedding dimensions of the measures are set as $(K, L, M) = (2, 3, 2)$.
As explained in the earlier sections, to calculate the measures we should first estimate the joint and single probability distributions from the obtained time series.
To get a reliable estimate within the limitation of the finite data set, it is preferable to keep the embedding dimensions relatively short \cite{MTE1}.

\subsection{Experiment 1}
\begin{figure}[htbp]
\centerline{\includegraphics[width=4.5in,bb=50 328 482 661, clip]{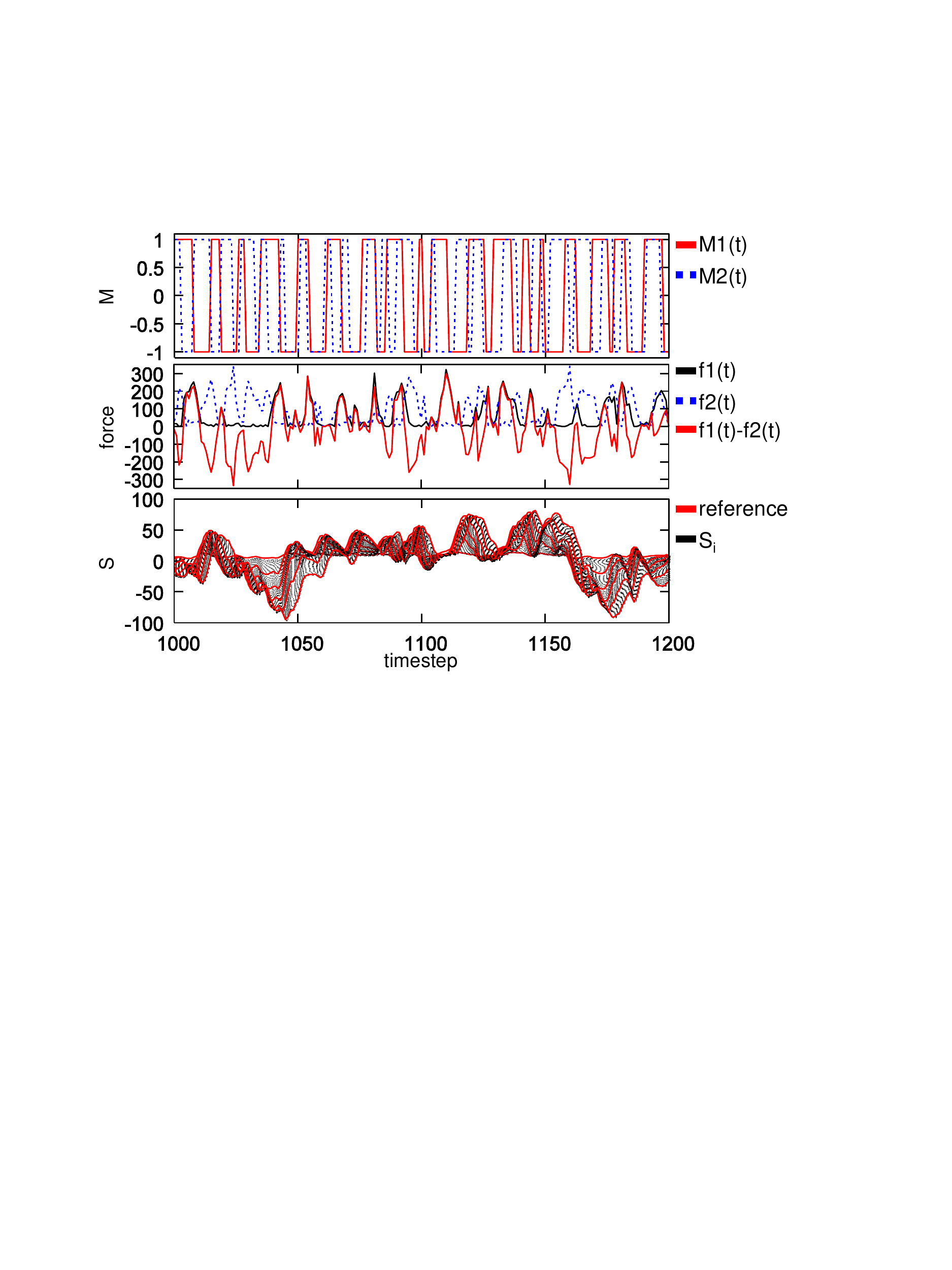}} %timeseries_exp1.eps
\caption{Typical examples of the time series for the random motor commands $(M1(t), M2(t))$ (upper plot), force sensors $(f1(t), f2(t), f1(t)-f2(t))$ (middle plot), and the body parts $S_{i}$ (lower plot) in Experiment 1. For the plots of the body parts, 101 time series (from $S_{1}$ to $S_{101}$) are overlaid and 6 reference tracking points are expressed in red lines.} \label{timeseries_exp1}
\end{figure}

In this experiment, we see how MSIT can quantitatively characterize the information-theoretic structure intrinsic to the soft robotic arm.
For this purpose, we adopted a random motor command for the actuation of the system to avoid additional correlations provided by the external. 
Figure \ref{timeseries_exp1} shows the example of the time series of the random motor commands $(M1(t), M2(t))$, the corresponding force sensory values $(f1(t), f2(t), f1(t)-f2(t))$, and the y-coordinates of the body parts $S_{i}$.
By using these obtained time series, we aim to analyze the following: (I) information transfers from the time series of the cable tension to that of each body part ($MSIT^{ST}_{f \rightarrow S_{i}} ( \tau )$) and (II) information transfers between the time series of each body part ($MSIT^{ST}_{S_{i} \rightarrow S_{j}} ( \tau )$).
In (I), since the cable tensions are driven by the motor command, this analysis is to see how the external actuation to the system transfers the information through the body.
For the analysis, we used the time series $f(t) = f1(t) - f2(t)$ for the force sensors, and calculated $MSIT^{ST}_{f \rightarrow S_{i}} ( \tau )$ for each body part $i$ and varying delay $\tau$.
In (II), we can characterize how one body part transfers information to another.
For the analysis, we calculated $MSIT^{ST}_{S_{i} \rightarrow S_{j}} ( \tau )$ for each pair of body parts $(i, j)$ and varying delay $\tau$. 
Note that we excluded the case when $i=j$.
For both analyses (I) and (II), $(N_{R}, T_{R})$ is set to $(1, 1)$ throughout the analysis and, to avoid the bias due to the finite data set, for each analysis of $MSIT^{ST}_{i \rightarrow j} ( \tau )$ we have also calculated the measure with temporally shuffled time series, and by iterating this procedure for 50 times and obtaining the averaged value, we have subtracted this value from the $MSIT^{ST}_{i \rightarrow j} ( \tau )$ in focus.     
For each experimental setting, we performed this procedure for 5 trials of experimental data and used the averaged $MSIT^{ST}_{i \rightarrow j} ( \tau )$ as a result.
In addition, when the information sender is within the range specified by $N_{R}$ and $T_{R}$, we need to take special care in defining $\hat{v}^{ST}_{i, j, t}$.
Details for this procedure are given in the Appendix.
Also, according to the setting of $N_{R}$ and $T_{R}$, we have excluded the case where the information receiver or sender is $S_{1}$ or $S_{101}$. 

\begin{figure}[htbp]
\centerline{\includegraphics[width=6.0in,bb=0 301 574 827, clip]{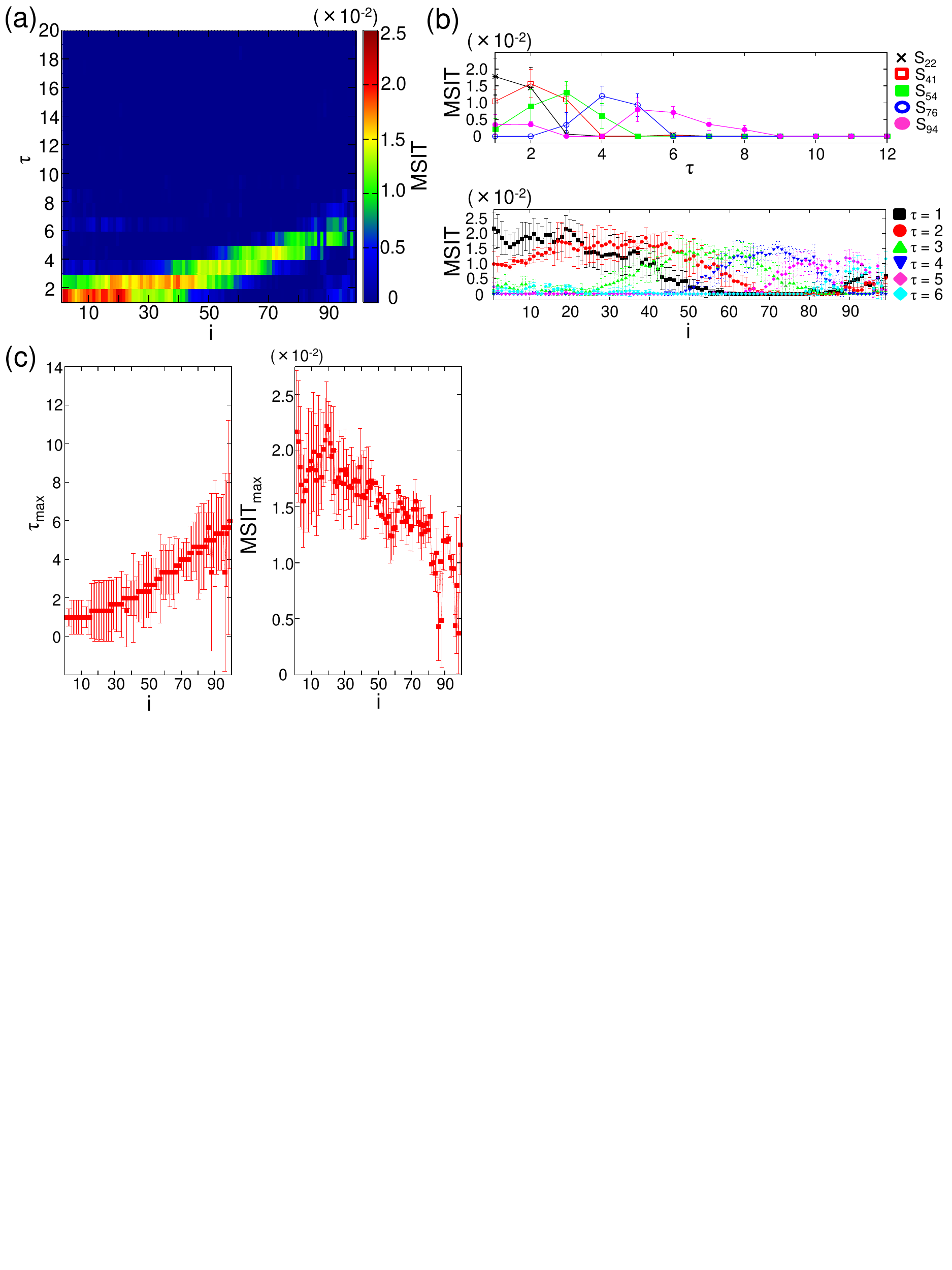}} %Motor_MSIT.eps
\caption{Results for MSIT from force sensors to the body parts. (a) Plot showing the averaged $MSIT^{ST}_{f \rightarrow S_{i}} ( \tau )$ in $i$-$\tau$ plane. (b) Plots overlaying several typical cross sections of (a) in $\tau$ axis (upper figure) and in $i$ axis (lower figure). (c) Plots showing $\tau_{max}$ (left figure) and $MSIT_{max}$ (right figure) according to each body part $i$. $\tau_{max}$ and $MSIT_{max}$ represent the $\tau$, where the MSIT is maximum, and the maximum MSIT over all delays $\tau$, respectively. For (b) and (c), the error bars show standard deviations.} \label{Motor_MSIT}
\end{figure}

Figure \ref{Motor_MSIT} (a) plots the results for the averaged $MSIT^{ST}_{f \rightarrow S_{i}} ( \tau )$ by varying the body part $i$ and the delay $\tau$.
We can clearly see that for body part closer to the base, more delayed MSIT is detected than for those closer to the tip.
Figure \ref{Motor_MSIT} (b) overlays several cross sections of Fig. \ref{Motor_MSIT} (a) in delay and space (body part).
As can be seen from the figure, in the body part near the arm base MSIT with shorter delay is dominant, while in the body part near the arm tip MSIT with relatively longer delay is dominant.
In addition, the strength of MSIT tends to decrease from the base to the tip.
To see this tendency in more detail, we have collected the maximum values of MSIT ($MSIT_{max}$) among delays and the delay $\tau_{max}$ that shows $MSIT_{max}$ for each body part and calculated the average for each (Fig. \ref{Motor_MSIT} (c)). 
As a result, we found that $\tau_{max}$ almost grows linearly as the body part changes from the base toward the tip and at around the tip $\tau_{max}$ took approximately 6 timesteps (Fig. \ref{Motor_MSIT} (c), left plot).
In addition, the corresponding $MSIT_{max}$ showed about $2.2 \times 10^{-2}$ bit at around the base ($S_{2}$), while the $MSIT_{max}$ showed a lower value of about $1.2 \times 10^{-2}$ bit at around the tip ($S_{100}$).  
Considering that 1 timestep is approximately 0.03 s in real physical time and the entire body is 310 mm long, this result implies that the dominant information transfer propagates with about 1.7 m/s in velocity through the body by losing the information for around $1.0 \times 10^{-2}$ bit.

\begin{figure}[htbp]
\centerline{\includegraphics[width=6.0in,bb=4 211 566 753, clip]{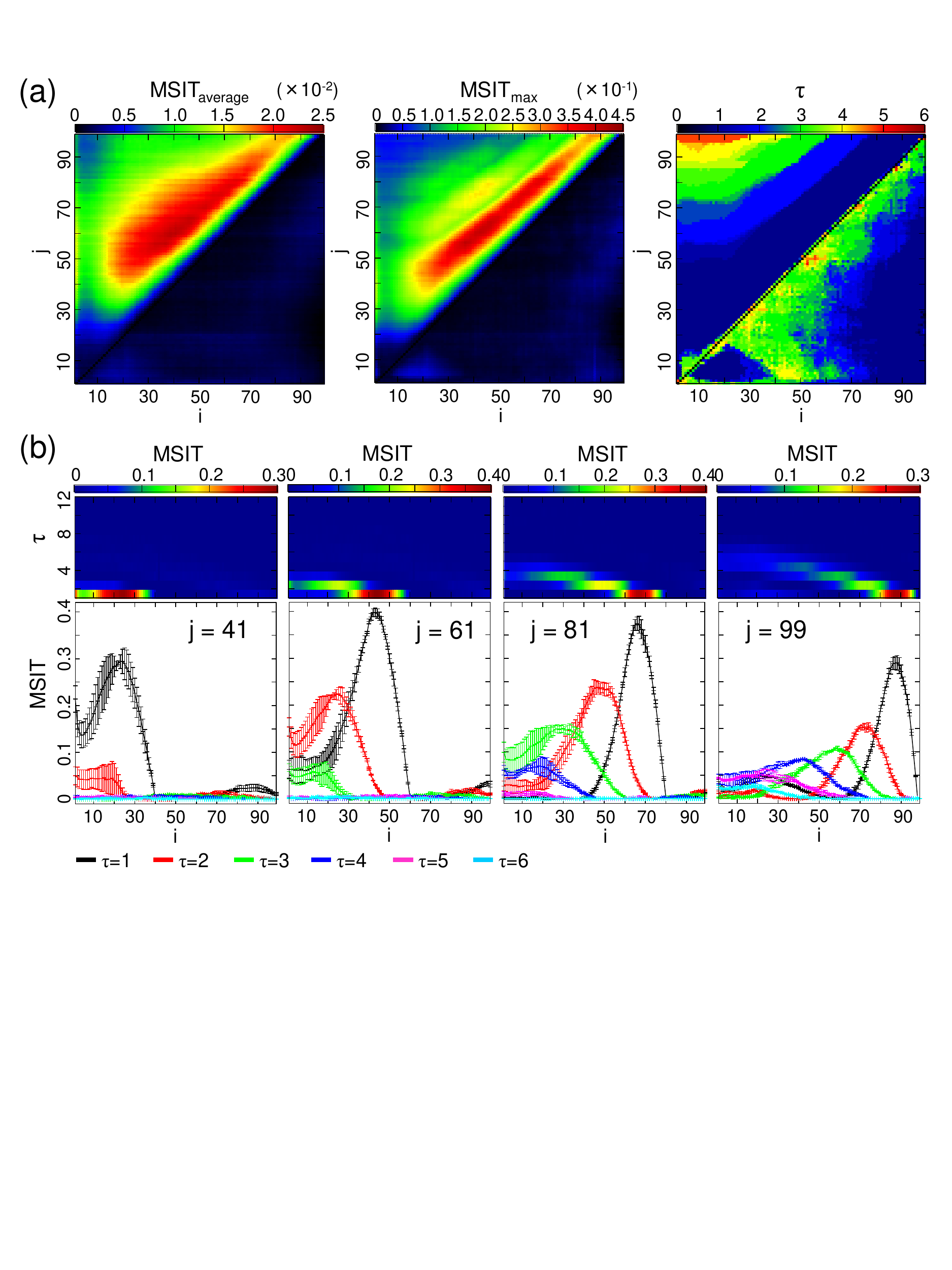}}% Body_Body_MSIT.eps
\caption{Results for MSIT between the body parts. (a) Plots showing the averaged $MSIT_{average}$ (left), $MSIT_{max}$ (middle), and $\tau_{max}$ (right) for each $(i, j)$ over 5 trials. See the main text for details. (b) Plots showing the averaged $MSIT^{ST}_{f \rightarrow S_{i}} ( \tau )$ over 5 trials according to each delay $\tau$ and the location of the information sender $i$ by fixing the location of the the information receiver to $j= 41, 61, 81,$ and $99$. The error bars show standard deviations.} \label{Body_Body_MSIT}
\end{figure}

Next, let us see how information transfers between each body part.
Here, for each information sender $S_{i}$ and receiver $S_{j}$ in the body parts, we calculated $MSIT^{ST}_{S_{i} \rightarrow S_{j}} ( \tau )$ by varying $\tau$ from 1 to 12 and obtained the averaged MSIT ($MSIT_{average}$) as $MSIT_{average} = \frac{1}{12} \sum_{\tau=1}^{12} MSIT^{ST}_{S_{i} \rightarrow S_{j}} ( \tau )$ for each trial. 
As in the previous experiments, we also collected the maximum MSIT ($MSIT_{max}$) among delays and the delay $\tau_{max}$ that takes $MSIT_{max}$ for each $(i, j)$ in each trial.
Figure \ref{Body_Body_MSIT} (a) shows the averaged $MSIT_{average}$, $MSIT_{max}$, and $\tau_{max}$ over 5 trials for each $(i, j)$.
In the plot for the $MSIT_{average}$ and $MSIT_{max}$, we can clearly see that high MSIT values are observed only in the region of $i < j$.
This suggests that the information is transferring in the direction from the base to the tip but not from the tip to the base.
This result is understandable when considering the experimental conditions of our platform, and is consistent with the results from the previous experiments.
Furthermore, from both plots we can confirm that, as the value for $j - i$ increases in the region of $i < j$, the values of $MSIT_{average}$ and $MSIT_{max}$ get gradually smaller.
This implies that, for each body part, information strongly transfers toward the nearest neighbor in the direction from the base to the tip.
For the results of $\tau_{max}$, large $\tau_{max}$ values are observed in the region for the smaller $i$ and the larger $j$.
This means that the information transfer takes longer time when the body parts are further away from each other.
It is noticeable that the largest $\tau_{max}$ in this region takes approximately 6 timesteps, which is also consistent with our previous results. 
To see these tendencies in further detail, by fixing the information receiver to several body parts we observed the behavior of $MSIT^{ST}_{S_{i} \rightarrow S_{j}} ( \tau )$ by varying the location of the information sender in the body $j$ and the delay $\tau$ (Fig. \ref{Body_Body_MSIT} (b)). 
We can clearly confirm from the plot that, for each information receiver, the largest information transfers with the shortest delay from neighboring body parts located in the base side (Fig. \ref{Body_Body_MSIT} (b)).
% Note that the value of $MSIT^{ST}_{S_{i} \rightarrow S_{j}} ( \tau )$ was generally larger than that of $MSIT^{ST}_{f \rightarrow S_{j}} ( \tau )$ in the previous experiment, which is also natural considering the spatial distance actuation

In this experiment, we have demonstrated how the measure MSIT can be effectively used to reveal the delayed information transfer in the soft robotic arm.
In the next experiment, we move on to characterize a local information transfer profile in the spatiotemporal dynamics of the soft body using LMSIT.

\subsection{Experiment 2}
Due to the soft flexible bodies, soft robots are sensitive to environmental/external stimulus in general. 
If the body receives some stimulus from the environment, the behavior of the body changes drastically and immediately.
This type of event often occurs at a specific point in time and space, which makes it difficult to detect and evaluate the effect of the stimulus using statistical methods based on the global average over the entire collected time series.  
In those cases, it would be beneficial if we could monitor what is happening locally at each spatiotemporal point in the dynamics.
In this experiment, we will see that LMSIT can be used effectively for this purpose.
LMSIT is defined for each information receiver and can measure the amount of information transferred from the sender at each spatiotemporal point.
Thus, the local information transfer profile in the spatiotemporal dynamics can be characterized.
As explained earlier, in this experiment, we drive the arm with the oscillatory motor command with an obstacle in the environment.  
We will see how the LMSIT ($LMSIT^{ST}_{S_{i} \rightarrow S_{j}} ( \tau )$) captures this effect from the environment on the soft robotic arm.
As a comparison, we will also drive the system without an obstacle in the environment with random motor commands and the same oscillatory motor command for 5000 timesteps and measure the LMSIT. 
In this experiment, we focus on the information transfer in the direction from the base to the tip (that is, $i < j$) by skipping the case when the time series of $S_{1}$ and $S_{101}$ becomes the information sender or receiver under consideration, taking into account the results of experiment 1.
In addition, similar to the setting of MSIT in experiment 1, $N_{R}$ and $T_{R}$ are both set to 1. 
If the information sender $S_{i}$ is in the neighboring point of $S_{j}$, we need to adjust the corresponding $\hat{v}^{ST}_{i, j, t-\tau}$.
Again, this procedure is described in the Appendix.

\begin{figure}[htbp]
\centerline{\includegraphics[width=6.0in,bb=3 187 577 784, clip]{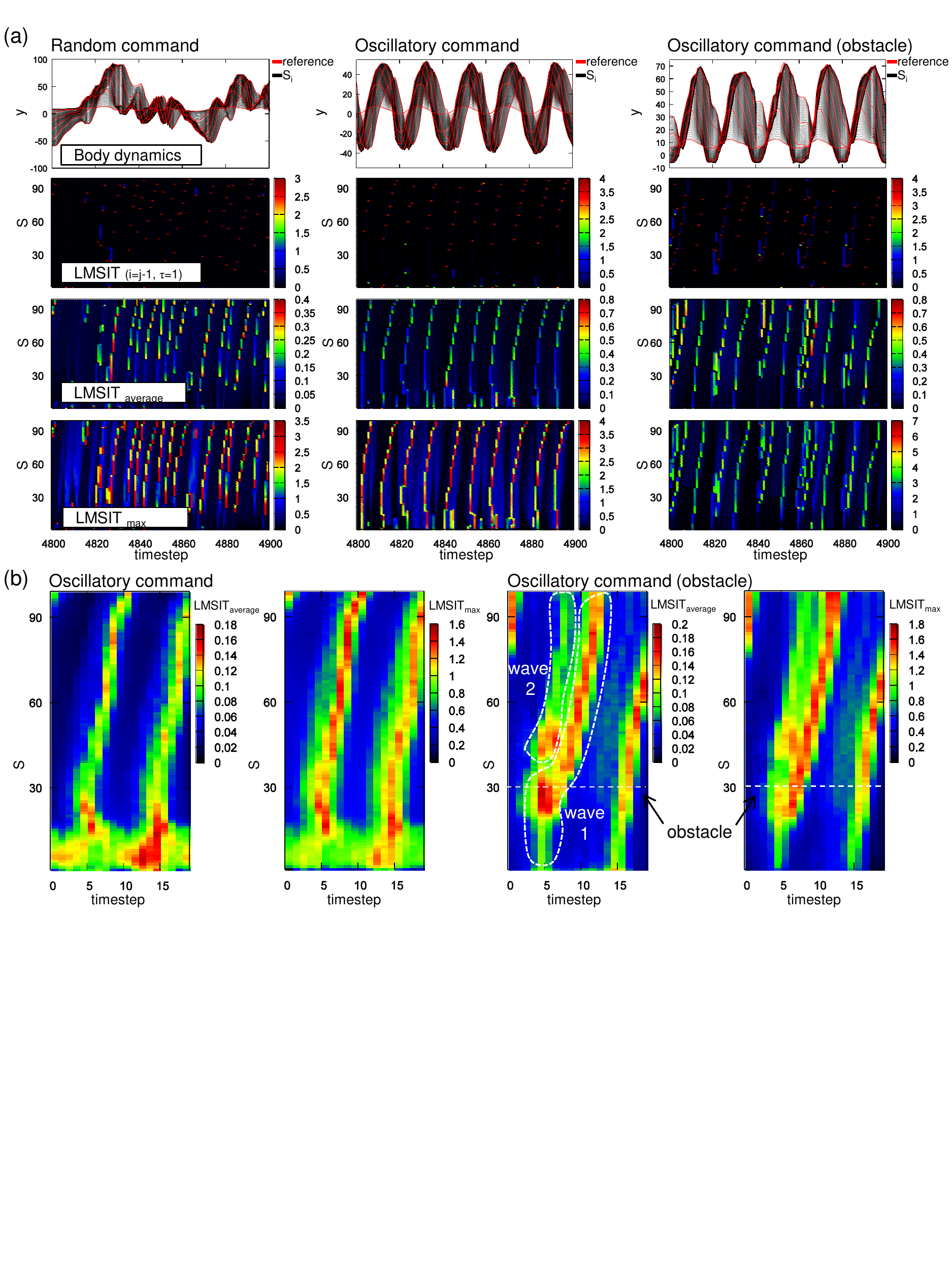}} %local_new.eps
\caption{Results for the LMSIT analyses. (a) Typical LMSIT profiles corresponding to the body dynamics for the case of the random motor command (left row), the oscillatory motor command (middle row), and the oscillatory motor command with an obstacle (right row). For each row, the overlaid plot for the $S_{i}$ time series, the corresponding LMSIT profile with $i=j-1$ and $\tau=1$, $LMSIT_{average}$ profile, and $LMSIT_{max}$ are presented from the top to the bottom. (b) The average $LMSIT_{average}$ and $LMSIT_{max}$ profile for the case with the oscillatory motor command without (left side) and with the obstacle (right side). Note that the timesteps when the motor commands switch to drive the arm toward the lower ($y<0$) and the upper ($y>0$) region in the screen correspond to timestep 5 and timestep 15, respectively, in both cases. The white line in the case with the obstacle represents the position of the contact point in relation to the obstacle. The white closures in the case with the obstacle depicts two spatially split information propagation waves, which we call {\it wave 1} and {\it wave 2}, respectively. See the main text for details.} \label{local}
\end{figure}

We start by observing the result for $LMSIT^{ST}_{S_{i} \rightarrow S_{j}} ( \tau )$ when $i=j-1$ and $\tau = 1$. 
Results are shown in the upper two plots of each row in Fig. \ref{local} (a). 
The first upper line shows the example of the time series in the case of the random motor command, the oscillatory motor command, and the same oscillatory motor command with the obstacle, from left to right.
The second upper line shows the corresponding $LMSIT^{ST}_{S_{i} \rightarrow S_{j}} ( \tau )$ with $i=j-1$ and $\tau = 1$.
In each plot, we observe high $LMSIT^{ST}_{S_{i} \rightarrow S_{j}} ( \tau )$ instantaneously in several spatiotemporal points.
However, the pattern of the information structure is not obvious.
To visualize the information transfer profile more clearly, by varying the delay $\tau$ from 1 to 20 and the information sender $S_{i}$ for all the body parts in $i < j$, we measured the average LMSIT ($LMSIT_{average}(j) = \frac{1}{20} \sum_{\tau=1}^{20} \frac{1}{j-1} \sum_{d=1, j-d < j}^{j-1} LMSIT^{ST}_{S_{j-d} \rightarrow S_{j}} ( \tau )$) and the maximum LMSIT ($LMSIT_{max}(j) = \max_{1 \leq \tau \leq 20, 1 \leq d \leq j-1} [LMSIT^{ST}_{S_{j-d} \rightarrow S_{j}} ( \tau )]$) for each information receiver $S_{j}$.
The results for $LMSIT_{average}(j)$ and $LMSIT_{max}(j)$ are shown in the lower two plots in Fig. \ref{local} (a).
In the case for the random motor command, we can clearly observe a high information propagation wave from the base toward the tip at the irregular temporal point.  
In the case of the oscillatory motor command, we observe a similar high information propagation wave from the base toward the tip two times in each oscillatory cycle at the specific timing.
In the case of the oscillatory motor command with the obstacle, we observe a clear high information propagation wave from the base toward the tip similar to the case with the oscillatory motor command without the obstacle in half of each oscillatory cycle, but when the arm hits the obstacle in the other half of each oscillatory cycle, we see the fluctuating information propagation wave, where sometimes the wave splits into two, from the base toward the tip. 

To see this fluctuating information propagation wave in detail, we averaged the $LMSIT_{average}(j)$ and $LMSIT_{max}(j)$ for each spatiotemporal point using 150 oscillation cycles for the oscillatory motor command with and without an obstacle (Fig. \ref{local} (b)).
As is obvious from the plots, in the case without the obstacle, we can clearly confirm two strong information propagation waves from the base toward the tip at specific timing, as we saw in Fig. \ref{local} (a).
In the case with the obstacle, we see that at the onset of the obstacle crush (around timestep 5), two spatially split strong information transfers are detected around the contact point of the object with the arm, and this information propagates toward the tip.
We speculate that one propagation wave from the base to the tip (wave 1) represents the propagation of the motor actuation through the body, and the other wave (wave 2) represents the fast shock wave due to the contact to the obstacle. 
Furthermore, we can see that the velocity of wave 1 is modulated at the contact point (Fig. \ref{local} (b)). 
This implies that the environmental damage provided by the crush to the obstacle induces a qualitatively different information transfer profile through the body from the one generated by the motor actuation.
We expect that LMSIT is capable of detecting the shock wave for the case of multiple-point contact as long as its effect is reflected in the body dynamics. 

\section{Discussion}
In this paper, by using the physical soft robotic arm platform, we demonstrated that the information-theoretic approach can be effective in characterizing the diverse spatiotemporal dynamics of soft bodies especially with delayed interaction.  
Currently, TE is one of the most applied measures to assess information transfers in robotics.
However, it has drawbacks in capturing delayed interaction and applying to multivariate time series.
In fact, these points are the major properties of soft body dynamics, which have high-dimensionality and memory.
To overcome these weak points, in this paper, we have introduced measures MSIT and LMSIT for applications to spatiotemporal dynamics of soft robots.
By using MSIT, we showed that the actuation to the arm transfers the information toward the tip with a specific time delay, and we evaluated this information transfer velocity.
Furthermore, we captured the delayed information transfer between each body part in detail.
By using LMSIT, we visualized the dynamic information transfer profile hidden in the spatiotemporal dynamics of the soft body, and characterized the environmental damage that spreads throughout the entire body.
These measures and the scheme can be applied effectively to soft robotic platforms in general.

Although we illustrated one possible scenario to characterize the information structure in a soft robotic platform, our approach can be extended in several ways taking into account each soft robotic platform in use and each experimental condition.
For example, in our experiment, we set the relation between space and time based on the sampling rate of the experimental devices.
However, the ratio between the experimentally observed space (the points in the body) and time does not always match the underlying physical dynamics, and this may make our proposed scheme ineffective. 
In those situations, experimenters should control the spatiotemporal scale of the experimentally obtained data.
In addition, as we saw in the definition of M(S)IT, that one important point to capture the delayed information transfer was to condition on other causal contributors.
Although we have introduced a reasonable physical assumption for conditioning, considering that soft robots can take unconventional morphology, it would be valuable in some situations to apply a method for reconstructing all the causal contributors from scratch only by using the obtained time series, which is introduced in \cite{MTE2,MTE3}.  
On the use of MIT, it has been recently reported in \cite{Wibral} that there is a case in which MIT does not effectively work in capturing the delayed couplings.
In \cite{Wibral}, an alternative measure called $TE_{SPO}$, where {\it SPO} stands for {\it self-prediction optimality}, has been proposed to measure the delayed information transfer.
Their approach can be also applied for the analyses.
Nevertheless, their measure is defined for the bivariate case. 
To extend the measure for the multivariate case, the concept presented in \cite{MTE2,MTE3} would be definitely useful and inevitable.

\section{Conclusion}
The behavior of animals or robots is generated by the dynamic coupling between the brain (controller), the body, and the environment.
These dynamic interactions between components are usually described based on separately defined dynamical systems (such as the neural network for the brain and physics and mechanics for the body). 
When illustrating a consistent view of the dynamic interaction modality among these components, several difficulties will arise, for example, in determining the time scales among the components and the threshold for modeling the physical contact.    
In this case, the information-theoretic approach can provide an effective method for giving a consistent view to characterize these dynamic couplings as information transfers, since the scheme is intrinsically based on a model-free approach.
In particular, it has been recently shown that soft bodies can be exploited as a computational resource (see, e.g., \cite{Kohei2,Kohei3}).
In this case, the soft body itself acts similarly to a huge recurrent neural network.
Although the brain has the apparent anatomical structure (network topology) to realize interactions among neurons, the body does not contain this type of explicit interaction pathway but has a characteristic morphology and can physically interact with the environment.
Our approach presented here would be also useful to reveal how each body part interacts with another, which is usually hidden in the dynamics, and to reconstruct a functional network topology within the body.
This type of analysis would be valuable to infer the information processing capacity of the body and is included in our future work.
Moreover, it would be valuable to investigate how the body morphology and the controller of soft robots self-organize in a given environment, guided by information transfers \cite{Martius}.
For example, by focusing on several variables in a soft robotic platform and by maximizing the information transfer between these variables, we can investigate how the design of the robot and the controller co-evolves \cite{Josh}, which would reveal the relations between the dynamic coupling and the resulting behavioral generation and help understanding a design principle for soft robots.

\ack
KN would like to acknowledge Dr. Tao Li for his help in collecting the experimental data and Dr. Hidenobu Sumioka for fruitful discussions in the early stage of the study.
KN, RF was supported by the European Commission in the ICT-FET OCTOPUS Integrating Project (EU project FP7-231608). 
KN was also supported by JSPS Postdoctoral Fellowships for Research Abroad.
NS, RF was supported by EU project Extending Sensorimotor Contingencies to Cognition (eSMCs), IST-270212.

\section*{Appendix}
\subsection*{Time series data preparation}
\begin{figure}[htbp]
\centerline{\includegraphics[width=5.5in,bb=3 362 573 779, clip]{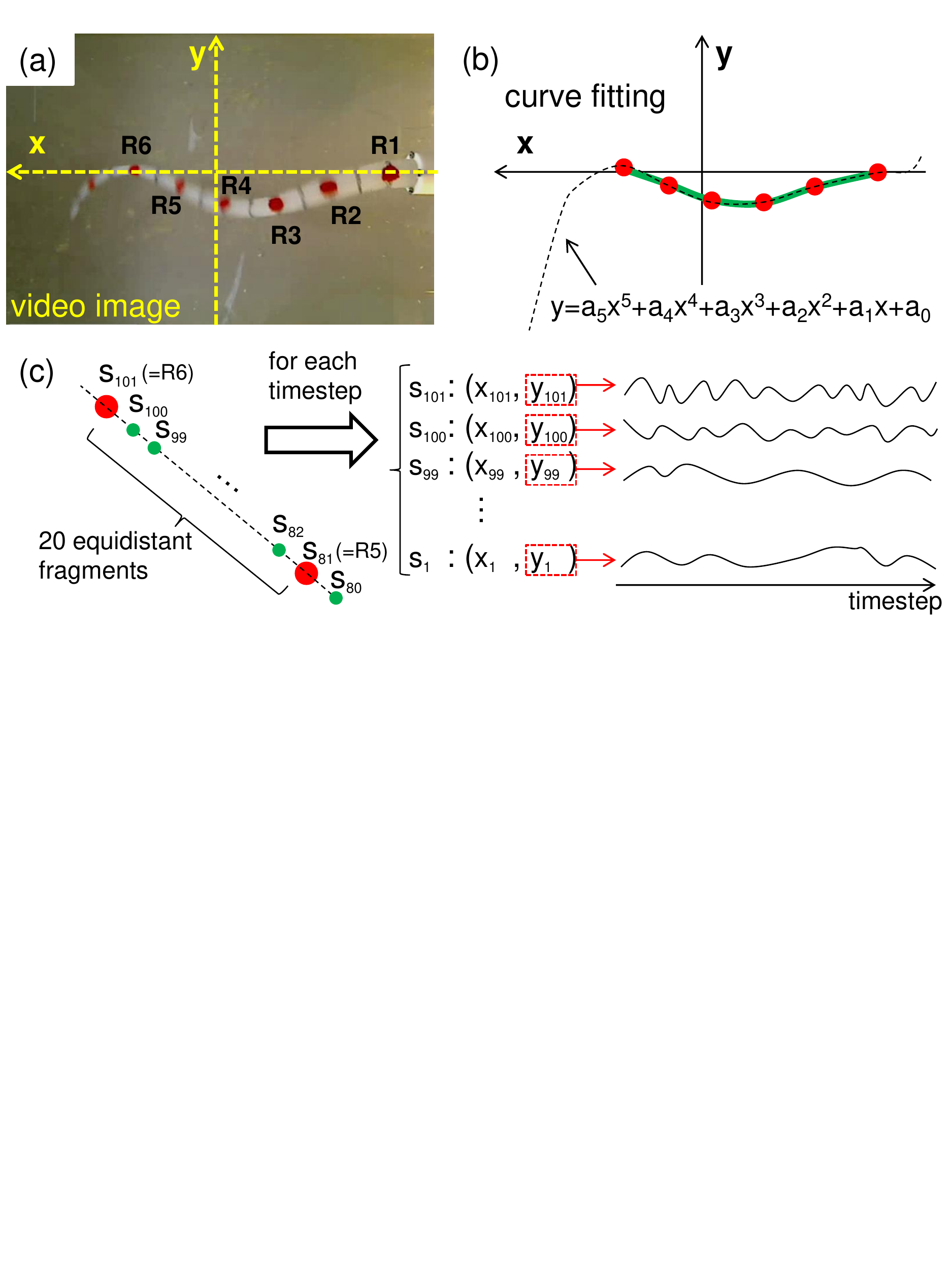}} %timeseries_preparation.eps
\caption{Schematics showing the procedures to prepare the time series representing the soft body dynamics. (a) For each timestep, the (x, y)-coordinates for 6 reference points are tracked using the tracking system \cite{Tracker}. Note that the nearest reference point to the tip is not used for this analysis (see the main text for details). (b) The 5th order polynomial curve fitting the body posture based on the obtained coordinates of the reference points. (c) Close-up of the curve between R6 and R5 as an example for the preparation of S. For each neighboring reference point, the curve is segmented into 20 equidistant fragments, and the endpoints of the fragments are indexed from $S_{1}$ to $S_{101}$ from the base to the tip, including the reference points, and their (x, y)-coordinates are collected for each timestep. The dynamics of the y-coordinates of these endpoints are the representation of the body dynamics in this paper.} \label{timeseries_preparation}
\end{figure}

%\begin{figure}[htbp]
%\centerline{\includegraphics[width=5.5in,bb=16 432 573 779, clip]{timeseries_preparation.eps}} 
%\caption{Schematics showing the procedures to prepare the time series representing the soft body dynamics. (a) For each timestep, the (x, y)-coordinates for 6 reference points are tracked using the tracking system \cite{Tracker}. Note that the nearest reference point to the tip is not used for this analysis (see the main text for details). (b) The 5th order polynomial curve fitting the body posture based on the obtained coordinates of the reference points. (c) Close-up of the curve between R6 and R5 as an example for the preparation of S. For each neighboring reference point, the curve is segmented into 20 equidistant fragments, and the endpoints of the fragments are indexed from $S_{1}$ to $S_{101}$ from the base to the tip, including the reference points, and their (x, y)-coordinates are collected for each timestep. The dynamics of the y-coordinates of these endpoints are the representation of the body dynamics in this paper.} \label{timeseries_preparation}
%\end{figure}
In this Appendix, we provide details on the time series preparation for S.
As explained in the main text, we first track the positions of 6 reference points in the body, $(x_{Ri}, y_{Ri})$ $(i=1, 2, ..., 6)$, for each timestep (Fig. \ref{timeseries_preparation} (a)) and fitted these points with a 5th order polynomial curve, expressed as
\begin{equation}
y=a_{5}x^{5} + a_{4}x^{4} + a_{3}x^{3} + a_{2}x^{2} + a_{1}x + a_{0},
\end{equation}  
where $a_{5}, a_{4}, a_{3}, a_{2}, a_{1},$ and $a_{0}$ are the constant parameters to be tuned (Fig. \ref{timeseries_preparation} (b)).
The choice of the order of a polynomial curve was determined by observing the arm motion, where the number of bends does not exceed what can be expressed as a 5th order polynomial curve.
Let us introduce a vector expression for these parameters, ${\bf A} = [a_{5}, a_{4}, a_{3}, a_{2}, a_{1}, a_{0}]^{\mathrm{T}}$.
By using the x-coordinates of the collected reference points, a matrix ${\bf X}$ is denoted as
\begin{eqnarray*}
{\bf X} = \left[
\begin{array}{cccccc}
x_{R1}^{5} & x_{R1}^{4} & x_{R1}^{3} & x_{R1}^{2} & x_{R1} & 1 \\
x_{R2}^{5} & x_{R2}^{4} & x_{R2}^{3} & x_{R2}^{2} & x_{R2} & 1 \\
x_{R3}^{5} & x_{R3}^{4} & x_{R3}^{3} & x_{R3}^{2} & x_{R3} & 1 \\
x_{R4}^{5} & x_{R4}^{4} & x_{R4}^{3} & x_{R4}^{2} & x_{R4} & 1 \\
x_{R5}^{5} & x_{R5}^{4} & x_{R5}^{3} & x_{R5}^{2} & x_{R5} & 1 \\
x_{R6}^{5} & x_{R6}^{4} & x_{R6}^{3} & x_{R6}^{2} & x_{R6} & 1 \\
\end{array}
\right].
\end{eqnarray*}
Then, denoting the y-coordinates of the collected reference points as ${\bf Y} = [y_{R1}, y_{R2}, y_{R3}, y_{R4}, y_{R5}, y_{R6}]^{\mathrm{T}}$,  
we can approximate the constant parameters using the Moore-Penrose pseudo-inverse as ${\bf A} = {\bf X}^{-1}{\bf Y}$.
Using the fitted curve, we fragmented the interval of each neighboring reference point into 20 equidistant segments along the curve (Fig. \ref{timeseries_preparation} (c)).
The resulting 101 endpoints for these segments are denoted from $S_{1}$ to $S_{101}$ from the base to the tip, including the reference points.
By iterating these procedures for each timestep, we collect y-coordinates for these endpoints resulting in 101 time series for $S_{1}, ..., S_{101}$ (Fig. \ref{timeseries_preparation} (c)).
These time series are the representation of the body dynamics and are used for the analyses.

\subsection*{Settings of possible information sources in our analyses}
\begin{figure}[htbp]
\centerline{\includegraphics[width=3.0in,bb=120 470 389 774, clip]{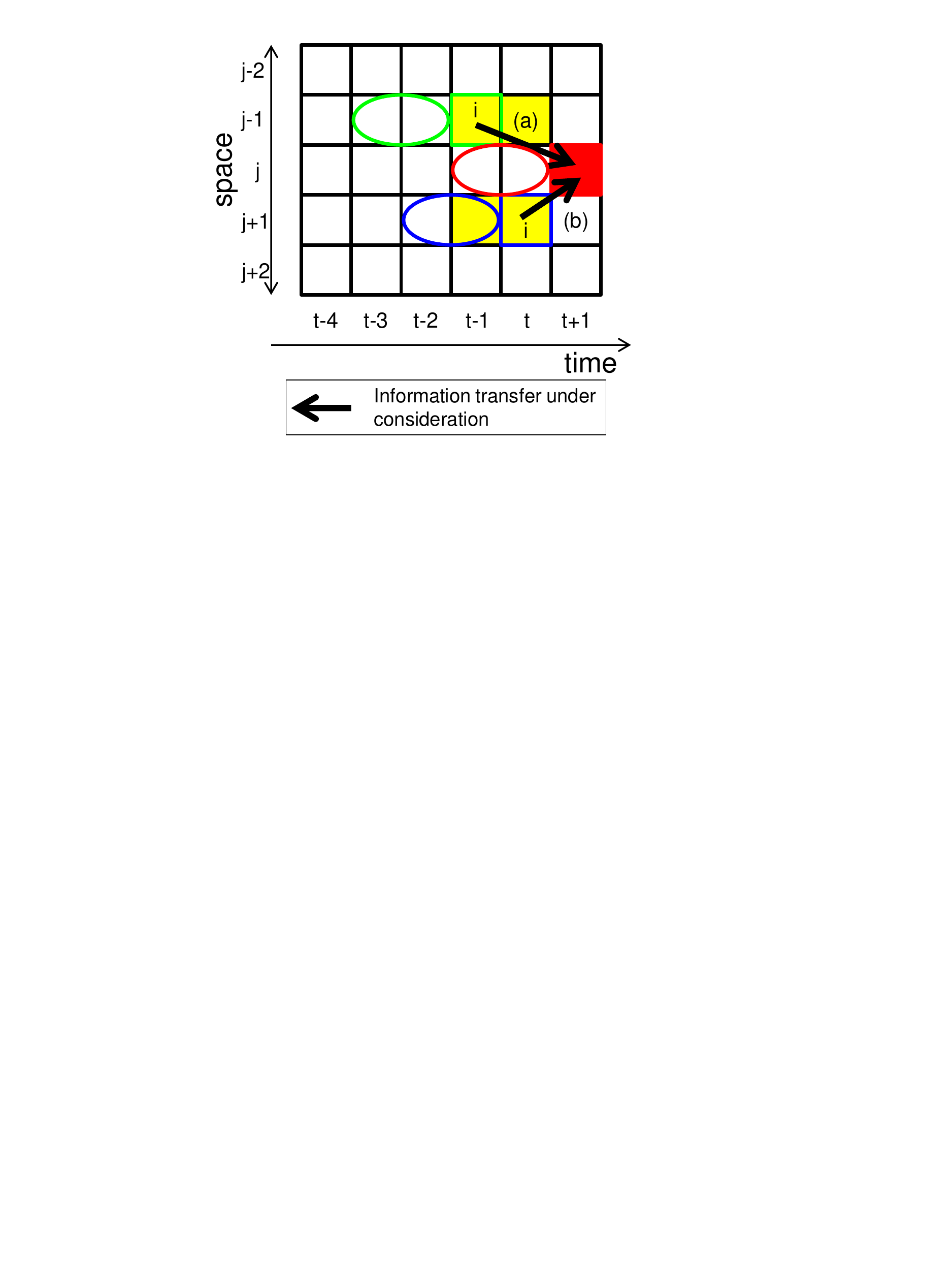}} %appendix_v.eps
\caption{Schematics explaining the setting of $\hat{v}^{ST}_{i, j, t-\tau}$ when the information sender $i$ is overlapped with $\hat{v}^{ST}_{i, j, t-\tau}$ through embedding dimensions. Here, $\hat{v}^{ST}_{i, j, t-\tau}$ is usually set as $\hat{v}^{ST}_{i, j, t-\tau} = \{ \hat{x}_{j+1, t}^{2}, \hat{x}_{j-1, t}^{2} \}$ from the definition. (a) Schematic expressing information transfer from the sender $i=j-1$ to the receiver $j$ with delay $\tau=2$. In this case, $\hat{v}^{ST}_{i, j, t-\tau}$ becomes to $\hat{v}^{ST}_{i, j, t-\tau} = \{ \hat{x}_{j+1, t}^{2} \}$. (b) Schematic expressing information transfer from the sender $i=j+1$ to the receiver $j$ with delay $\tau=1$. In this case, $\hat{v}^{ST}_{i, j, t-\tau}$ becomes to $\hat{v}^{ST}_{i, j, t-\tau} = \{ \hat{x}_{j-1, t}^{2} \}$. For each information sender and receiver, the circle expresses the condition under its own past.
Note that, in the figure, $(K, L, M, N_{r}, T_{r}) = (2, 3, 2, 1, 1)$.} \label{appendix_v}
\end{figure}

Here, we explain how we set the condition of possible information sources $\hat{v}^{ST}_{i, j, t-\tau}$ ($\hat{v}^{ST}_{i, j, t}$) in our analyses.
For descriptive purposes, we unify the expression of possible information sources as $\hat{v}^{ST}_{i, j, t-\tau}$ in this section.
In our analyses, if the information sender has an overlap with $\hat{v}^{ST}_{i, j, t-\tau}$ through embedding dimensions, then we discarded the term containing the overlap in $\hat{v}^{ST}_{i, j, t-\tau}$.
Typical cases for this setting are exemplified in Fig. \ref{appendix_v}.

\clearpage
\clearpage
\clearpage
\clearpage
\clearpage
\clearpage
\clearpage
\clearpage


\section*{References}
\begin{thebibliography}{10}
%Intro
\bibitem{Soft1} Pfeifer R, Lungarella M and Iida F 2012 The challenges ahead for bio-inspired `soft' robotics {\it Commun. ACM.} {\bf 55} 76-87
\bibitem{Soft2} Trivedi D, Rahn CD, Kier WM and Walker ID 2008 Soft robotics: biological inspiration, state of the art, and future research {\it Appl. Bionics. Biomech.} {\bf 5} 99-117
\bibitem{Soft3} Kim S, Laschi C and Trimmer B 2013 Soft robotics: a new perspective in robot evolution {\it Trends Biotechnol.} {\bf 31} 287-94
\bibitem{Soft4} Shepherd RF, Ilievski F, Choi W, Morin SA, Stokes AA, Mazzeo AD, Chen X, Wang M and Whitesides GM 2011 Multi-gait soft robot {\it Proc. Natl. Acad. Sci. U. S. A.} {\bf 108} 20400-3
\bibitem{Rolf1} Pfeifer R, Lungarella M and Iida F 2007 Self-organization, embodiment, and biologically inspired robotics {\it Science} {\bf 318} 1088-93
\bibitem{Rolf2} Pfeifer R and Bongard J 2006 {\it How the Body Shapes the Way We Think: A New View of Intelligence} (MIT Press, Cambridge, MA)
\bibitem{Kohei1} Nakajima K, Ngouabeu AMT, Miyashita S, G\"{o}ldi M, F\"{u}chslin RM and Pfeifer R 2012 Morphology-Induced Collective Behaviors: Dynamic Pattern Formation in Water-Floating Elements {\it PLoS ONE} {\bf 7} e37805
\bibitem{Oct1} Hochner B 2012 An embodied view of octopus neurobiology {\it Curr. Biol.} {\bf 22} R887-92
\bibitem{Oct3} Sumbre G, Fiorito G, Flash T and Hochner B 2005 Motor control of flexible octopus arms {\it Nature} {\bf 433} 595-6
\bibitem{Oct4} Sumbre G, Gutfreund Y, Fiorito G, Flash T and Hochner B 2001 Control of octopus arm extension by a peripheral motor program {\it Science} {\bf 293} 1845-8
\bibitem{Soft_SSSA3} Laschi C, Mazzolai B, Mattoli V, Cianchetti M and Dario P 2009 Design of a biomimetic robotic octopus arm {\it Bioinsp. Biomim.} {\bf 4} 015006
\bibitem{Soft_SSSA1} Cianchetti M, Arienti A, Follador M, Mazzolai B, Dario P and Laschi C 2011 Design concept and validation of a robotic arm inspired by the octopus {\it Mater. Sci. Eng. C} {\bf 31} 1230-9
\bibitem{Soft_SSSA4} Laschi C, Mazzolai B, Cianchetti M, Margheri L, Follador M and Dario P 2012 A soft robot arm inspired by the octopus {\it Adv. Robot.} {\bf 26} 709-27
\bibitem{Soft_SSSA2} Calisti M, Giorelli M, Levy G, Mazzolai B, Hochner B, Laschi C and Dario P 2011 An octopus-bioinspired solution to movement and manipulation for soft robots {\it Bioinsp. Biomim.} {\bf 6} 036002
\bibitem{Tao2} Li T, Nakajima K, Calisti M, Laschi C and Pfeifer R 2012 Octopus-Inspired Sensorimotor Control of a Multi-Arm Soft Robot {\it Proc. Int. Conf. on Mechatronics and Automation} (Chengdu, China, 5-8 August) pp 948-55
\bibitem{Oct2} Li T, Nakajima K, Kuba MJ, Gutnick T, Hochner B and Pfeifer R 2011 From the octopus to soft robots control: an octopus inspired behavior control architecture for soft robots {\it Vie et Milieu} {\bf 61} 211-7
\bibitem{Kuwabara} Kuwabara J, Nakajima K, Kang R, Branson DT, Guglielmino E, Caldwell DG and Pfeifer R 2012 Timing-Based Control via Echo State Network for Soft Robotic Arm {\it Proc. Int. Joint Conf. on Neural Networks} (Brisbane, Australia, 10-15 June) pp 1-8
\bibitem{Tao1} Li T, Nakajima K and Pfeifer R 2013 Online Learning Technique for Behavior Switching in a Soft Robotic Arm {\it Proc. IEEE Int. Conf. on Robotics and Automation} (Karlsruhe, Germany, 6-10 May) pp 1296-1302
\bibitem{Tao3} Li T, Nakajima K, Cianchetti M, Laschi C and Pfeifer R 2012 Behavior Switching by Using Reservoir Computing for a Soft Robotic Arm {\it Proc. IEEE Int. Con. on Robotics and Automation} (Saint Paul, MN, USA, 14-18 May) pp 4918-24
\bibitem{Kohei2} Nakajima K, Hauser H, Kang R, Guglielmino E, Caldwell DG and Pfeifer R 2013 A soft body as a reservoir: case studies in a dynamic model of octopus-inspired soft robotic arm {\it Front. Comput. Neurosci.} {\bf 7} 1-19
\bibitem{Kohei3} Nakajima K, Hauser H, Kang R, Guglielmino E, Caldwell DG and Pfeifer R 2013 Computing with a Muscular-Hydrostat System {\it Proc. IEEE Int. Conf. on Robotics and Automation} (Karlsruhe, Germany, 6-10 May) pp 1504-11
\bibitem{Amblard} Amblard PO and Michel OJJ 2011 On directed information theory and Granger causality graphs {\it J. Comput. Neurosci.} {\bf 30} 7-16
\bibitem{Kantz} Marschinski R and Kantz H 2002 Analysing the information flow between financial time series - an improved estimator for transfer entropy {\it Eur. Phys. J. B} {\bf 30} 275-281
\bibitem{MTE1} Pompe B and Runge J 2011 Momentary information transfer as a coupling measure of time series {\it Phys. Rev. E} {\bf 83} e051122
\bibitem{MTE2} Runge J, Heitzig J, Petoukhov V and Kurths J 2012 Escaping the Curse of Dimensionality in Estimating Multivariate Transfer Entropy {\it Phys. Rev. Lett.} {\bf 108} 258701
\bibitem{MTE3} Runge J, Heitzig J, Marwan N and Kurths J 2012 Quantifying causal coupling strength: A lag-specific measure for multivariate time series related to transfer entropy {\it Phys. Rev. E} {\bf 86} 061121
\bibitem{TE} Schreiber T 2000 Measuring information transfer {\it Phys. Rev. Lett.} {\bf 85} 461-4
\bibitem{Nico} Schmidt N, Hoffmann M, Nakajima K and Pfeifer R 2013 Bootstrapping Perception Using Information Theory: Case Studies in a Quadruped Robot Running on Different Grounds {\it Adv. Complex Syst.} {\bf 16} 1250078
\bibitem{Max} Lungarella M and Sporns O 2006 Mapping information flow in sensorimotor networks {\it PLoS Comput. Biol.} {\bf 2} 1301-12
%Information
\bibitem{Cover} Cover TM and Thomas JA 1991 {\it Elements of Information Theory} (Wiley-Interscience)
\bibitem{Wibral} Wibral M, Pampu N, Priesemann V, Siebenh\"{u}hner F, Seiwert H, Lindner M, Lizier JT and Vicente R 2013 Measuring Information-Transfer Delays {\it PLoS ONE} {\bf 8} e55809
\bibitem{Polani} Ay N and Polani D 2008 Information Flows in Causal Networks {\it Adv. Complex Syst.} {\bf 11} 17-41
\bibitem{Lizier2} Lizier JT and Prokopenko M 2010 Differentiating information transfer and causal effect {\it Eur. Phys. J. B} {\bf 73} 605-615
\bibitem{PE1} Bandt C and Pompe B 2002 Permutation entropy: A natural complexity measure for time series {\it Phys. Rev. Lett.} {\bf 88} e174102
\bibitem{STE} Staniek M and Lehnertz K 2008 Symbolic transfer entropy {\it Phys. Rev. Lett.} {\bf 100} e158101.
\bibitem{PE3} Bahraminasab A, Ghasemi F, Stefanovska A, McClintock PVE and Kantz H 2008 Direction of coupling from phases of interacting oscillators: A permutation information approach {\it Phys. Rev. Lett.} {\bf 100} e084101
\bibitem{PE5} Rosso OA, Larrondo HA, Martin MT, Plastino A and Fuentes MA 2007 Distinguishing noise from chaos {\it Phys. Rev. Lett.} {\bf 99} e154102
\bibitem{TERV2} Kugiumtzis D 2013 Partial transfer entropy on rank vectors {\it Eur. Phys. J. Special Topics} {\bf 222} 401-20.
\bibitem{Kohei4} Nakajima K and Haruna T 2013 Symbolic local information transfer {\it Eur. Phys. J. Special Topics} {\bf 222} 421-39
\bibitem{PE4} Cao YH, Tung WW, Gao JB, Protopopescu VA and Hively LM 2004 Detecting dynamical changes in time series using the permutation entropy {\it Phys. Rev. E} {\bf 70} e046217
\bibitem{PE2} Bandt C, Keller G and Pompe B 2002 Entropy of interval maps via permutations {\it Nonlinearity} {\bf 15} 1595-1602
\bibitem{Amigo_book} Amig\'{o} JM 2010 {\it Permutation Complexity in Dynamical Systems} (Springer-Verlag: Berlin, Heidelberg, Germany)
\bibitem{Amigo1} Amig\'{o} JM, Kennel MB, Kocarev L 2005 The permutation entropy rate equals the metric entropy rate for ergodic information sources and ergodic dynamical systems {\it Physica D} {\bf 210} 77-95
\bibitem{Amigo3} Amig\'{o} JM, Keller K 2013 Permutation entropy: One concept, two approaches {\it Eur. Phys. J. Special Topics} {\bf 222} 263-73
\bibitem{Haruna1} Haruna T and Nakajima K 2011 Permutation complexity via duality between values and orderings {\it Physica D} {\bf 240} 1370-7
\bibitem{Amigo2} Amig\'{o} JM 2012 The equality of Kolmogorov-Sinai entropy and metric permutation entropy generalized {\it Physica D} {\bf 241} 789-93
\bibitem{Keller1} Keller K, Unakafov AM and Unakafova VA 2012 On the relation of KS entropy and permutation entropy {\it Physica D} {\bf 241} 1477-81
\bibitem{Keller2} Unakafova VA, Unakafov AM and Keller K 2013 An approach to comparing Kolmogorov-Sinai and permutation entropy {\it Eur. Phys. J. Special Topics} {\bf 222} 353-61
\bibitem{Keller3} Keller K and Sinn M 2009 A standardized approach to the Kolmogorov-Sinai entropy {\it Nonlinearity} {\bf 22} 2417-22
\bibitem{Keller4} Keller K and Sinn M 2010 Kolmogorov-Sinai entropy from the ordinal viewpoint {\it Physica D} {\bf 239} 997-1000
\bibitem{Keller5} Keller K 2012 Permutations and the Kolmogorov-Sinai entropy {\it Discr. Cont. Dyn. Syst.} {\bf 32} 891-900
\bibitem{Haruna2} Haruna T and Nakajima K 2013 Permutation Complexity and Coupling Measures in Hidden Markov Models {\it Entropy} {\bf 15} 3910-30
\bibitem{Haruna3} Haruna T and Nakajima K 2013 Permutation approach to finite-alphabet stationary stochastic processes based on the duality between values and orderings {\it Eur. Phys. J. Special Topics} {\bf 222} 367-83
\bibitem{Haruna4} Haruna T and Nakajima K 2013 Symbolic transfer entropy rate is equal to transfer entropy rate for bivariate finite-alphabet stationary ergodic Markov processes {\it Eur. Phys. J. B} {\bf 86} 230
\bibitem{TERV1} Kugiumtzis D 2012 Transfer entropy on rank vectors {\it J. Nonlin. Sys. Appl.} {\bf 3} 73-81.
\bibitem{Lizier1} Lizier JT, Prokopenko M, Zomaya AY 2008 Local information transfer as a spatiotemporal filter for complex systems {\it Phys. Rev. E} {\bf 77} 026110
\bibitem{Robio2} Nakajima K, Li T, Kang R, Guglielmino E, Caldwell DG and Pfeifer R 2012 Local Information Transfer in Soft Robotic Arm {\it Proc. IEEE Int. Conf. on Robotics and Biomimetics} (Guangzhou, China, 11-14 December) pp 1273-80
\bibitem{Tracker} Brown D, Tracker video analysis and modeling tool. http://www.cabrillo.edu/ dbrown/tracker/, December, 2009.
\bibitem{Robio1} Nakajima K, Li T, Sumioka H, Cianchetti M and Pfeifer R 2011 Information Theoretic Analysis on a Soft Robotic Arm Inspired by the Octopus {\it Proc. IEEE Int. Con. on Robotics and Biomimetics} (Phuket, Thailand, 7-11 December) pp 110-7
\bibitem{Martius} Martius G, Der R and Ay N 2013 Information driven self-organization of complex robotic behaviors. {\it PLoS ONE} {\bf 8} e63400
\bibitem{Josh} Bongard J 2011 Morphological change in machines accelerates the evolution of robust behavior. {\it Proc. Natl. Acad. Sci. U. S. A.} {\bf 108} 1234-9
\end{thebibliography}
\end{document}